\documentclass[sigplan,nonacm,screen]{acmart}

\AtBeginDocument{%
  }

\usepackage[]{hyperref}
\usepackage{mathtools}
\usepackage{braket}
\usepackage{physics}
\usepackage[normalem]{ulem}
\usepackage{bbm}
\usepackage{graphicx}
\usepackage{tikz}
\usetikzlibrary{fit,positioning,calc,arrows.meta,backgrounds}
\usepackage{quantikz}

\definecolor{qsblue}{HTML}{ACD5EB}
\definecolor{qsred}{HTML}{E49CA3}
\definecolor{qsgreen}{HTML}{B3D5AA}
\definecolor{qspurple}{HTML}{BEAECF}
\definecolor{qscream}{HTML}{FCE0C4}
\definecolor{qsorange}{HTML}{F8B168}
\definecolor{qsink}{HTML}{333333}
\definecolor{qsgrid}{HTML}{DDDDDD}
\usepackage{subcaption}
\usepackage{siunitx}
\usepackage{enumitem}
\usepackage{float}
\usepackage{amsmath, algorithm, algpseudocode}
\DeclareRobustCommand*\circled[1]{\tikz[baseline=(char.base)]{
            \node[shape=circle,draw,inner sep=1pt] (char) {#1};}}

\newcommand{\quasistim}{\textsc{QuasiStim}}
\newcommand{\quasifold}{\textsc{QuasiFold}}
\newcommand{\deltaframe}{\textsc{DeltaFrame}}
\newcommand{\mcsampler}{\texttt{MCSampler}}

\begin{document}
\title{Scalable Exact Non-Pauli Noise Simulation for Fault-Tolerant Quantum Circuits}

\author{Yuewen Hou}
\email{isaachyw@umich.edu}
\orcid{0009-0004-3035-3197}
\affiliation{%
  \institution{University of Michigan}
  \city{Ann Arbor}
  \state{MI}
  \country{USA}
}

\author{Gokul Subramanian Ravi}
\email{gsravi@umich.edu}
\orcid{0000-0002-2334-2682}
\affiliation{%
  \institution{University of Michigan}
  \city{Ann Arbor}
  \state{MI}
  \country{USA}
}

\begin{abstract}

Fault-tolerant quantum design relies on simulating noisy quantum error correction (QEC) circuits with hundreds to thousands of physical qubits and large circuit depths. High-throughput stabilizer simulators, such as Stim, are the de facto tools for this regime. However, these simulators typically support only Clifford gates and stochastic Pauli noise. As a result, more realistic device noise models—such as coherent control errors and thermal relaxation—are often approximated using Pauli twirling. This replacement offers speed but can introduce inaccuracies.

To address this, we propose \quasistim{}\footnote{\quasistim{} will be open-sourced upon publication at \url{https://github.com/isaachyw/QuasiStim}.} for scalable exact QEC simulation of
realistic noise, with \quasifold{} as a compiler frontend and \deltaframe{} as
a sampling backend. \quasifold{} lowers sampling overhead by merging equivalent
non-Pauli sites in the quantum circuit. \deltaframe{} samples the resulting stabilizer trajectories by
reusing precomputed predicted references instead of resimulating the full
tableau for each shot.
\quasifold{} can reduce the number of shots by over $100\times$ on bivariate-bicycle (BB) codes.
Compared to a per-shot tableau simulator, \deltaframe{}
is $166\times$--$1331\times$ faster on rotated surface-code memory under
coherent $R_Z$ noise as $d$ grows from 5 to 35. \quasistim{} also makes the PTA
bias measurable: the gap is non-negligible and varies with noise strength,
decoder, and circuit schedule.

\end{abstract}


\maketitle 

\section{Introduction}\label{sec:introduction}
\newsavebox{\introcircuitbox}
\sbox{\introcircuitbox}{%
  \begin{quantikz}[font=\scriptsize, row sep={0.58cm,between origins},
                   column sep=0.18cm, thin lines]
    & \gate{H} & \ctrl{1} & \gate{M} & \\
    &          & \targ{}  & \gate{M} &
  \end{quantikz}}

\begin{figure}[t]
\centering
\scalebox{1.2}{%
\begin{tikzpicture}[
    font=\scriptsize,
    panel/.style={font=\footnotesize\bfseries, anchor=west},
    spine/.style={draw=qsink, line width=0.55pt},
    gline/.style={draw=qsgrid, line width=0.4pt},
    bar/.style={draw=qsink, line width=0.6pt},
    pbox/.style={draw=qsink, line width=0.6pt, rounded corners=2pt},
    flow/.style={-latex, draw=qsink, line width=0.7pt},
    ytick/.style={anchor=east, inner sep=1.5pt},
    xtick/.style={anchor=north, inner sep=3pt},
  ]

  \node[panel] at (0.05,-0.18) {(a) Sampling time};
  \begin{scope}[shift={(1.45,-2.18)}]
    \foreach \y in {0.652,1.304} \draw[gline] (0,\y) -- (6.6,\y);
    \draw[spine] (0,0) -- (6.6,0);
    \draw[spine] (0,0) -- (0,1.63);
    \foreach \y/\t in {0/0, 0.652/40, 1.304/80} {
      \draw[spine] (-0.07,\y) -- (0,\y);
      \node[ytick] at (-0.09,\y) {\t};
    }
    \node[rotate=90] at (-0.74,0.815) {Time ($\mu$s/shot)};

    \draw[bar, fill=qsblue]  (0.93,0) rectangle (2.67,1.175);
    \draw[bar, fill=qsgreen] (3.93,0) rectangle (5.67,0.10);
    \node[anchor=south] at (1.80,1.20) {$72.06$};
    \node[anchor=south] at (4.80,0.13) {$0.243$};
    \node at (4.80,0.98) {$297\times$ speedup};
    \node[xtick] at (1.80,-0.02) {Tableau baseline};
    \node[xtick] at (4.80,-0.02) {\quasistim{}};
  \end{scope}

  \node[panel] at (0.05,-2.92) {(b) Logical error rate gap};
  \begin{scope}[shift={(1.45,-4.95)}]
    \foreach \y in {0.516,1.031,1.547} \draw[gline] (0,\y) -- (6.6,\y);
    \draw[spine] (0,0) -- (6.6,0);
    \draw[spine] (0,0) -- (0,1.65);
    \foreach \y/\t in {0/4, 0.516/5, 1.031/6, 1.547/7} {
      \draw[spine] (-0.07,\y) -- (0,\y);
      \node[ytick] at (-0.09,\y) {\t};
    }
    \node[rotate=90] at (-0.74,0.825) {LER ($\times 10^{-4}$)};

    \draw[bar, fill=qsblue]  (0.93,0) rectangle (2.67,1.084);
    \draw[bar, fill=qsgreen] (3.93,0) rectangle (5.67,0.596);
    \node[anchor=north] at (1.80,1.055) {$6.10$};
    \node[anchor=north] at (4.80,0.567) {$5.16$};
    \draw[spine] (1.80,1.202) -- (1.80,1.253) -- (4.80,1.253) -- (4.80,1.202);
    \node[anchor=south] at (3.30,1.271) {PTA $+18.4\%$};
    \node[xtick] at (1.80,-0.02) {PTA};
    \node[xtick] at (4.80,-0.02) {\quasistim{}};

    \draw[white, line width=3pt] (0,-0.075) -- (0,0.075);
    \foreach \y in {-0.045,0.045}
      \draw[spine] (-0.086,\y-0.030) -- (0.086,\y+0.030);
  \end{scope}

  \node[panel] at (0.05,-5.66) {(c) Circuit + device noise};
  \begin{scope}[shift={(0,-8.30)}]
    \node[pbox, fill=white, anchor=south west,
          minimum width=3.72cm, minimum height=1.58cm] (qec) at (0.16,0.80) {};
    \node[anchor=north] at (2.02,2.34) {QEC circuit};
    \node at (2.02,1.38) {\usebox{\introcircuitbox}};

    \node[pbox, fill=qscream, anchor=south west,
          minimum width=3.72cm, minimum height=0.63cm] (dev) at (0.16,0.05) {};
    \node[anchor=center] at (2.02,0.50) {Realistic device noise};
    \node[anchor=center, font=\tiny] at (2.02,0.21) {rotation / relaxation};

    \node[pbox, fill=qsgreen, minimum width=3.14cm, minimum height=0.61cm]
         (qs) at (6.32,1.49) {\footnotesize\bfseries \quasistim{}};

    \draw[flow] (3.96,1.49) -- (4.71,1.49);
    \draw[flow] (3.96,0.36) -- (4.71,1.23);
    \draw[flow] (6.32,1.16) -- (6.32,0.74);
    \node[anchor=center] at (6.32,0.50) {Logical error rate};
  \end{scope}

\end{tikzpicture}%
}
\caption{\quasistim{} samples the specified non-Pauli channel at Pauli-frame
  speed, without replacing it by its Pauli twirl.  Panels (a) and (b) use
  distance-5 rotated surface $X$-memory under thermal relaxation, on different
  circuit schedules.
  (a) \deltaframe{} samples one trajectory in $0.243\,\mu$s against
  $72.06\,\mu$s for a per-shot tableau baseline, a $297\times$ speedup.
  (b) PTA overestimates the logical error rate by $18.4\%$, reporting
  $6.10\times10^{-4}$ against the exact $5.16\times10^{-4}$.
  (c) The \quasistim{} flow: a QEC circuit and a realistic device-noise model in, decoded
  logical error rate out.}
  \label{fig:intro-overview}
\end{figure}
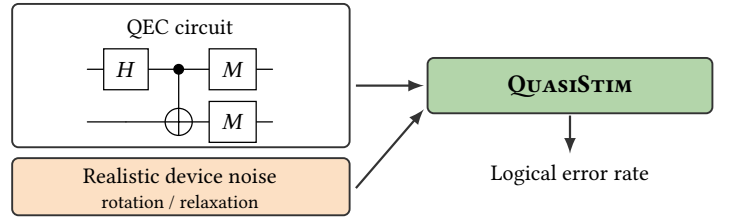

Quantum computers offer the prospect of solving problems for which no efficient
classical algorithm is known. Realizing that prospect at useful problem sizes
requires logical error rates far below those of present physical qubits, which
are achievable only in a fault-tolerant architecture
\cite{Cain_et_al_2026,Webster_et_al_2026}. Experimental systems have now
reached the early stages of that regime \cite{GoogleQuantumAI_2025,Bluvstein_et_al_2026}.
Quantum error-correction (QEC) codes, syndrome-extraction circuits, and decoders
are consequently no longer purely theoretical constructs; they are design
objects whose cost and performance must be evaluated quantitatively to maximize execution fidelity and efficiency.

Physical quantum hardware at the required scale does not yet exist. A complete
fault-tolerant algorithm would occupy thousands to millions of physical qubits,
so research on QEC, decoders, and related topics is conducted mostly via classical
simulation today. The objective of classical simulation of quantum systems varies as quantum computing matures. In
the NISQ regime, a simulator evaluates a complete program, expressed as a
logical circuit, on tens of qubits. In the fault-tolerant regime, the assembled
algorithm is beyond the reach of any simulator, so simulation is instead
performed on the physical circuits of individual ``gadgets'': quantum memory,
lattice surgery, and magic-state factories. Resource-estimation frameworks
subsequently compose the measured gadget error rates into an
architecture-level cost
\cite{Campbell_et_al_2026,Kan_Harkness_Du_Rofougaran_Garner_Liu_Mao_Stein_2026}.

Three prominent quantum simulators serve this process with different goals. Stim~\cite{Gidney_2021} samples Clifford
circuits under stochastic Pauli noise at millions of shots per second, which is
the throughput that QEC studies require at low logical error rates. Clifft and Tsim~\cite{Chase_Labib_2026,Haenel_Luo_Zhao_2026}
extend exact simulation to circuits containing a small number of $T$ gates, at a
memory and time cost that grows exponentially in the non-Clifford resource. All
three, however, share one assumption: the noise itself is a stochastic Pauli
channel. Clifft and Tsim spend their non-Clifford budget on the circuit, in gadgets such
as noisy magic-state cultivation or distillation, rather than on the noise
model.

Real device error characteristics, however, do not satisfy that assumption. Superconducting,
trapped-ion, and neutral-atom platforms exhibit coherent control errors, thermal
relaxation, amplitude damping, leakage, and qubit loss
\cite{Hashim_et_al_2021,Miao_et_al_2023,Wu_Kolkowitz_Puri_Thompson_2022}. These noise
channels produce correlated and coherent error paths that a Pauli mixture cannot accurately
reproduce, and they change, for instance, which syndromes a decoder observes
\cite{Zhou_Ji_Ding_2025,Takou_Brown_2025,Garner_Myers_Wang_Stein_Liu_Li_2025}.
The standard remedy is the Pauli twirling approximation (PTA), which replaces
each channel by its ``twirl''~\cite{Geller_Zhou_2013}. The twirl is a stochastic
Pauli channel, so Stim's frame sampler remains applicable, recovering high-throughput QEC simulation.
However, while PTA preserves the process fidelity of each individual channel, it is not faithful to the channel itself: by discarding the coherence that the noise imparts to the state, it biases the fidelity of the full circuit. Thus, without a scalable simulator for the original noise
channel, an error-resilient protocol or a decoder cannot be thoroughly evaluated against
the noise the device actually produces.
This is especially detrimental to long-term fault-tolerant quantum computing (FTQC) studies, where biased gadget
error rates are extrapolated across distances and composed into crucial resource
estimates.

In this work, we present \quasistim{} to close that gap. The user supplies a QEC
circuit and a device-noise model, and \quasistim{} returns decoded logical error
rates without substituting a twirl for the specified noise channel. The design
retains Stim's central idea. A stabilizer circuit can be simulated by tracking
how it transforms Pauli operators using a bookkeeping structure called a tableau, but
that simulation is the expensive step. Stim pays this cost once, on the noiseless
circuit, to obtain a reference measurement record, and then recovers each noisy
shot's record by tracking how its errors flip bits relative to that reference.
Quasiprobability decomposition (QPD) makes the same split possible for non-Pauli
noise. Each channel is written as a signed combination of stabilizer channels,
so a shot draws one of them at every noise site --- we call each such choice a
branch --- and the circuit it executes remains a stabilizer circuit. Averaged
over draws, the estimator is unbiased
\cite{Bennink_Ferragut_Humble_Laska_Nutaro_Pleszkoch_Pooser_2017}.

Exactness is not free and incurs two major costs. The number of shots needed grows as
$\Gamma^2$, where $\Gamma$ is the product over sites of how much negative weight
each channel's decomposition carries. Furthermore, because the branches a shot draws
change the circuit it executes, shots no longer share a single reference record, so a
conventional sampler simulates a fresh tableau for every shot. \quasistim{}
attacks these two costs separately. \quasifold{} is a compiler that rewrites the noise
sites to make $\Gamma$ smaller, merging rotations about a common axis, removing
rotations no measurement can detect, and absorbing Pauli noise into a
neighboring site where beneficial (\S\ref{sec:quasifold}). \deltaframe{} is a
sampler that works out the reference record for each branch ahead of time, so
that shots which drew different branches can still be advanced together as Pauli
frames (\S\ref{sec:deltaframe-compile}).

Figure~\ref{fig:intro-overview} shows the framework's example benefits on a distance-5
rotated surface $X$-memory under thermal relaxation. \deltaframe{} samples one
trajectory in $0.243\,\mu$s against $72.06\,\mu$s for a per-shot tableau
baseline, a $297\times$ speedup. On the same code and noise model, PTA
overestimates the logical error rate by $18.4\%$, reporting
$6.10\times10^{-4}$ where the exact channel gives $5.16\times10^{-4}$. These two results are complementary. The first establishes that exact simulation is affordable at QEC scale; the second establishes that it is necessary, since the twirl misestimates the logical error rate at an operating point of practical interest.

Both effects hold across our broad evaluations. \quasifold{} reduces the required shot
count by $1.16\times$ to $38.3\times$ on the surface code and over $100\times$ on BB code memory
(\S\ref{sec:eval-onenorm}). \deltaframe{} is $166\times$--$1331\times$
faster than a per-shot tableau sampler on the surface code under coherent $R_Z$
noise as $d$ grows from 5 to 35 (\S\ref{sec:eval-cost-model}). With \quasistim{}'s scalable high-faithfulness execution
capability, we are able to measure the PTA--exact gap systematically on QEC primitives spanning memory, logical operations, and magic-state cultivation
(\S\ref{sec:eval-gap}). As our results show, the gap is not determined by noise strength alone. Its sign and magnitude depend on the decoder, circuit schedule, code, and the balance of Pauli and non-Pauli noise. This variability demonstrates that exact simulation is necessary: the twirling gap cannot be reliably predicted in advance.
Finally, we note that while this framework is generally suitable for diverse and realistic noise, tailoring for leakage and qubit loss channels is out of scope and left to future work.

Overall, we make the following contributions:
\begin{itemize}
\item \quasistim{}, a QPD-based simulator that samples general non-Pauli noise on
  QEC-scale Clifford circuits while retaining Stim's throughput.
\item \quasifold{}, a channel-level compiler that fuses equivalent CPTP sites,
  prunes unobservable ones, and folds Pauli noise when the site one-norm falls, reducing the required shot count by 1--2 orders of magnitude on different codes and code distances.
\item \deltaframe{}, a sampling backend that precomputes branch-dependent
  references and executes mixed QPD patterns as Pauli frames, which is 2--3 orders of magnitude faster than a per-shot tableau sampler on different codes and distances, under
  coherent $R_Z$ noise, at a measured accuracy cost
  quantified in \S\ref{sec:eval-accuracy}.
\item A PTA-versus-exact LER study across memory, logical operations, and
  magic-state cultivation, showing that the twirling gap depends on the decoder,
  the schedule, the code, and the noise mix.
\end{itemize}

\section{Background}
\label{sec:background}

In this section, we first describe Stim, the method of choice for simulating noisy
Clifford QEC circuits under stochastic Pauli noise, and the frame-based sampling
that makes it fast.  We then describe stabilizer-channel QPD, which represents a non-Pauli completely positive and
trace-preserving (CPTP) channel exactly in expectation while keeping every
sampled trajectory as a stabilizer circuit, so that one shot costs one
polynomial-time tableau simulation.  We close with the Pauli twirling
approximation (PTA), the standard way to retain Stim's fast path, and show by
example what it discards in order to do so.

\subsection{Stim's Fast Simulation}

An $n$-qubit stabilizer state is represented by a generating set of Pauli
operators instead of a $2^n$-dimensional state vector.
Figure~\ref{fig:tableau} gives such a generating set, written in Pauli terms,
for the circuit of Figure~\ref{fig:stim-frames}.  Clifford gates map Pauli
operators to Pauli operators under conjugation, which enables polynomial-time
simulation of Clifford circuits \cite{Aaronson_Gottesman_2004}.  A dense
stabilizer tableau stores the binary $X$ and $Z$ components of these generators
in $O(n^2)$ bits.

Stim performs the tableau simulation \textbf{only once} to produce a noiseless
reference record~\cite{Gidney_2021}.  Noisy shots are then sampled as Pauli
frames: each frame is an $O(n)$-bit Pauli string, and XORing it with the
reference outcomes captures the effect of Pauli noise on that shot.  A Pauli
operation updates a frame in $O(1)$ time, and because frames are algebraically
simple, bulk updates are bit-packed and executed with SIMD instructions.
Figure~\ref{fig:stim-frames} illustrates this process.  The result is the high-throughput simulation that circuit-level QEC studies at low logical error rates
require, but that throughput is available only for Pauli noise: a general
non-Pauli channel has no frame representation, so it falls outside this approach entirely.

\begin{figure*}[t]
\centering
\scalebox{1.2}{%
\begin{tikzpicture}[
    font=\footnotesize,
    panel/.style={draw, dashed, rounded corners=2pt, inner sep=4pt},
    op/.style={draw, fill=white, minimum height=4.2mm, minimum width=5.6mm,
               inner xsep=1.5pt, inner ysep=0pt, font=\scriptsize},
    noise/.style={op, dashed, fill=qsorange!55},
    targ/.style={draw, circle, fill=white, minimum size=3.0mm, inner sep=0pt},
    cell/.style={draw, fill=white, minimum size=3.6mm, inner sep=0pt,
                 font=\scriptsize},
    hot/.style={cell, fill=qsorange!45},
    wire/.style={semithick},
    guide/.style={densely dotted, black!45},
    head/.style={font=\scriptsize\bfseries, anchor=west},
    ann/.style={font=\scriptsize, anchor=west},
    cost/.style={font=\scriptsize\itshape, anchor=west},
    flow/.style={-latex, thick},
  ]

  \begin{scope}
    \foreach \y in {0.62,0.0,-0.62} { \draw[wire] (-0.02,\y) -- (4.05,\y); }
    \node[font=\scriptsize] (la) at (-0.32,0.62)  {$a$};
    \node[font=\scriptsize]      at (-0.32,0.0)   {$d_1$};
    \node[font=\scriptsize] (ld) at (-0.32,-0.62) {$d_2$};

    \foreach \y in {0.62,0.0,-0.62} {
      \node[op] at (0.30,\y) {$R$};
      \node[op] at (0.98,\y) {$H$};
    }
    \draw[wire] (1.66,0.62) -- (1.66,0.0);
    \draw[wire] (2.34,0.62) -- (2.34,-0.62);
    \fill (1.66,0.62) circle (1.5pt);
    \fill (2.34,0.62) circle (1.5pt);
    \node[targ] at (1.66,0.0)   {};
    \node[targ] at (2.34,-0.62) {};
    \foreach \x/\y in {1.66/0.0, 2.34/-0.62} {
      \draw[semithick] (\x-0.11,\y) -- (\x+0.11,\y);
      \draw[semithick] (\x,\y-0.11) -- (\x,\y+0.11);
    }
    \node[op] at (3.02,0.62) {$H$};
    \node[op] (meas) at (3.70,0.62) {$M$};

    \draw[flow] (2.00,-0.98) -- (2.00,-1.28);
    \node[font=\scriptsize, anchor=east] at (1.78,-1.50) {$r_0=$};
    \node[cell, fill=qsblue!45] (r0c) at (1.98,-1.50) {$0$};
    \node[ann] (na) at (-0.50,-1.95)
         {noiseless record, $r_0\in\mathbb{F}_2^{L}$ ($L{=}1$ here)};

    \node[panel, draw=qsblue!85, fit=(la)(ld)(meas)(r0c)(na)] (pa) {};
    \node[head] at (-0.50,1.58) {\circled{1} reference: one tableau run};
    \node[cost, anchor=north west] at ($(pa.south west)+(0,-0.13)$)
         {tableau cost $O(ng+nm_{\mathrm{d}}+n^2m_{\mathrm{r}})$, paid once};
  \end{scope}

  \draw[flow] (4.25,0.10) -- (4.80,0.10);

  \begin{scope}[xshift=5.44cm]
    \foreach \y in {0.62,0.0,-0.62} { \draw[wire] (-0.02,\y) -- (4.73,\y); }
    \node[font=\scriptsize] (lb) at (-0.32,0.62)  {$a$};
    \node[font=\scriptsize]      at (-0.32,0.0)   {$d_1$};
    \node[font=\scriptsize]      at (-0.32,-0.62) {$d_2$};

    \foreach \y in {0.62,0.0,-0.62} {
      \node[op] at (0.30,\y) {$R$};
      \node[op] at (0.98,\y) {$H$};
    }
    \node[noise] (err) at (1.66,0.0) {$Z$};
    \draw[wire] (2.34,0.62) -- (2.34,0.0);
    \draw[wire] (3.02,0.62) -- (3.02,-0.62);
    \fill (2.34,0.62) circle (1.5pt);
    \fill (3.02,0.62) circle (1.5pt);
    \node[targ] at (2.34,0.0)   {};
    \node[targ] at (3.02,-0.62) {};
    \foreach \x/\y in {2.34/0.0, 3.02/-0.62} {
      \draw[semithick] (\x-0.11,\y) -- (\x+0.11,\y);
      \draw[semithick] (\x,\y-0.11) -- (\x,\y+0.11);
    }
    \node[op] at (3.70,0.62) {$H$};
    \node[op] at (4.38,0.62) {$M$};

    \foreach \x in {1.66,2.34,3.02,3.70} {
      \draw[guide] (\x,-0.86) -- (\x,-1.08);
    }
    \node[ann] (fh) at (-0.50,-0.94) {frame $(x,z)$};
    \foreach \y/\q in {-1.34/a, -1.72/{d_1}, -2.10/{d_2}} {
      \node[font=\scriptsize] at (0.30,\y) {$\q$};
    }
    \foreach \y in {-1.34,-1.72,-2.10} { \node[cell] at (0.98,\y) {$I$}; }
    \node[cell] at (1.66,-1.34) {$I$};
    \node[hot]  at (1.66,-1.72) {$Z$};
    \node[cell] at (1.66,-2.10) {$I$};
    \node[hot]  at (2.34,-1.34) {$Z$};
    \node[cell] at (2.34,-1.72) {$Z$};
    \node[cell] at (2.34,-2.10) {$I$};
    \node[cell] at (3.02,-1.34) {$Z$};
    \node[cell] at (3.02,-1.72) {$Z$};
    \node[cell] at (3.02,-2.10) {$I$};
    \node[hot]  at (3.70,-1.34) {$X$};
    \node[cell] at (3.70,-1.72) {$Z$};
    \node[cell] (lastc) at (3.70,-2.10) {$I$};

    \node[font=\scriptsize] at (4.38,-0.99) {$f_s$};
    \node[hot] (fs) at (4.38,-1.34) {$1$};

    \node[ann] (nb) at (-0.50,-2.48)
         {$X$ on $a$ anticommutes with $M_Z$: this record bit flips};

    \node[panel, draw=qsgreen!85, fit=(lb)(fh)(err)(fs)(lastc)(nb)] (pb) {};
    \node[head] at (-0.50,1.58) {\circled{2} Pauli frame: one per shot};
    \node[cost, anchor=north west] at ($(pb.south west)+(0,-0.13)$)
         {$O(n)$ bits, $O(1)$ per gate; no tableau};
  \end{scope}

  \draw[flow] (10.37,0.10) -- (10.92,0.10);

  \begin{scope}[xshift=11.01cm]
    \node[font=\scriptsize] (wh) at (1.95,0.90) {one word per frame bit};
    \foreach \y/\lb in {0.55/{x_a}, 0.15/{z_a}, -0.25/{z_{d_1}}} {
      \node[font=\scriptsize, anchor=west] at (0.05,\y) {$\lb$};
    }
    \node[font=\scriptsize] (vd) at (0.30,-0.56) {$\vdots$};

    \foreach \x in {0.75,1.15,1.55,2.35,2.75,3.15} {
      \node[cell] at (\x,0.55)  {$0$};
      \node[cell] at (\x,0.15)  {$0$};
      \node[cell] at (\x,-0.25) {$0$};
    }
    \node[hot] at (1.95,0.55)  {$1$};
    \node[hot] at (1.95,0.15)  {$0$};
    \node[hot] (jc) at (1.95,-0.25) {$1$};
    \node[cell] at (1.15,0.15) {$1$};
    \node[cell] at (2.75,0.55) {$1$};
    \node[cell] (lastw) at (3.15,-0.25) {$1$};
    \node[font=\scriptsize] (sj) at (1.95,-0.62) {shot $j$ shaded};

    \node[ann] (nx) at (0.05,-1.18) {one CNOT $=$ two word XORs};

    \node[draw, rounded corners=1.5pt, fill=qsred!35, font=\scriptsize,
          inner xsep=3pt, minimum height=5.0mm] (comb) at (1.85,-1.8)
         {\circled{4}\ $\mathrm{rec}_s=r_0\oplus f_s$};
    \node[ann] (nc) at (0.05,-2.40)
         {$r_0$ from \circled{1}, $f_s$ from \circled{2}};

    \node[panel, draw=qspurple!85, fit=(wh)(vd)(sj)(jc)(lastw)(nx)(comb)(nc)] (pc) {};
    \node[head] at (0.05,1.58) {\circled{3} SIMD bit-slicing};
    \node[cost, anchor=north west] at ($(pc.south west)+(0,-0.13)$)
         {$W_{\mathrm{SIMD}}$ shots per word operation};
  \end{scope}

\end{tikzpicture}%
}
\caption{Stim pays the tableau cost once per circuit and reduces each
shot to linear frame algebra.  The example circuit is one $X$-type
parity check.
\circled{1} One noiseless run is simulated with the inverse Clifford tableau
and its measurement record is kept as $r_0$; this is the only step that pays
the tableau cost, $O(ng+nm_{\mathrm{d}}+n^2m_{\mathrm{r}})$ for $n$ qubits, $g$
gates, $m_{\mathrm{d}}$ deterministic measurements, and $m_{\mathrm{r}}$ random measurements.  \circled{2} A
shot inserts Pauli noise, here a $Z$ on $d_1$, and propagates it as a Pauli frame:
the first CNOT copies the $Z$ from its target onto the ancilla, the second
leaves that $Z$ unchanged, and the
closing $H$ maps that $Z$ to $X$, which anticommutes with $M_Z$ and flips the
record bit.  The frame is two bits per qubit and a fixed-size Clifford or Pauli
updates it in $O(1)$ time, so no tableau is touched after \circled{1}.
\circled{3} Frames are stored bit-sliced, with bit $j$ of each machine word
holding shot $j$, so one word-wide XOR advances $W_{\mathrm{SIMD}}$ shots at
once, where $W_{\mathrm{SIMD}}$ is the SIMD register width.  \circled{4} The record of shot $s$ is then
$r_0\oplus f_s$.  The split holds only when the per-shot perturbation is a
Pauli: a non-Pauli site has no single frame to propagate, and
\S\ref{sec:deltaframe-compile} extends the same reference/frame split to
heterogeneous quasiprobability branch patterns.
}
\label{fig:stim-frames}
\end{figure*}
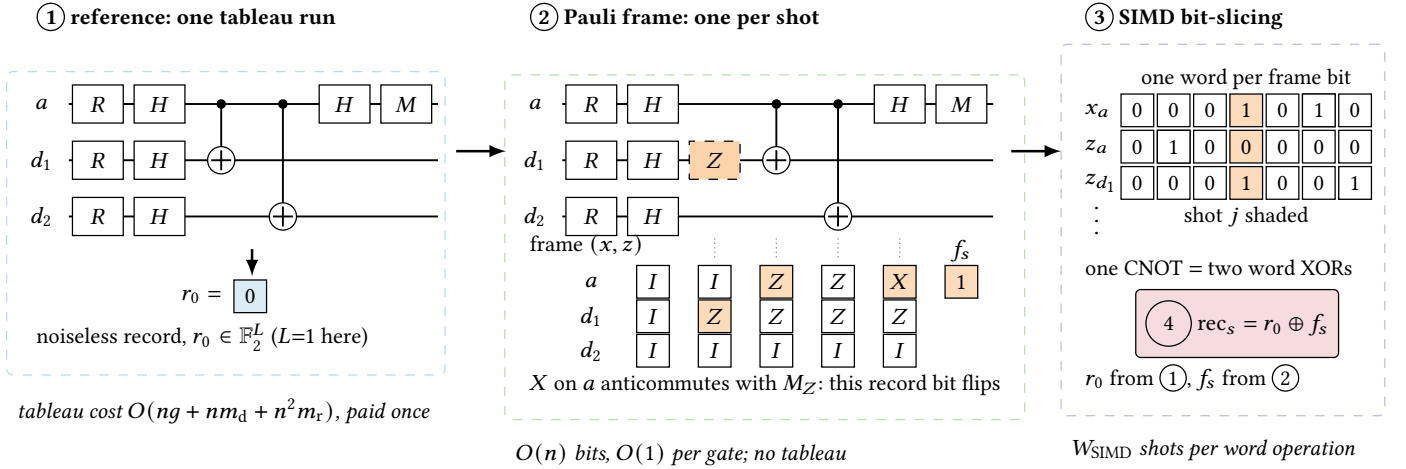

\begin{figure}[t]
\centering
\scalebox{1.2}{%
\begin{tikzpicture}[
    font=\small,
    hdr/.style={font=\small\bfseries},
    ann/.style={font=\small, anchor=west},
    cell/.style={font=\small},
  ]

  \node[hdr] at (0.00,0.95) {};
  \node[hdr] at (0.95,0.95) {$a$};
  \node[hdr] at (1.70,0.95) {$d_1$};
  \node[hdr] at (2.45,0.95) {$d_2$};
  \node[hdr, anchor=west] at (3.05,0.95) {sign};
  \draw[semithick] (-0.55,0.72) -- (4.10,0.72);

  \node[ann] at (-0.55,0.40) {$g_1$};
  \node[cell] at (0.95,0.40) {$Z$};
  \node[cell] at (1.70,0.40) {$X$};
  \node[cell] at (2.45,0.40) {$X$};
  \node[cell, anchor=west] at (3.05,0.40) {$+$};

  \node[ann] at (-0.55,0.02) {$g_2$};
  \node[cell] at (0.95,0.02) {$I$};
  \node[cell] at (1.70,0.02) {$X$};
  \node[cell] at (2.45,0.02) {$I$};
  \node[cell, anchor=west] at (3.05,0.02) {$+$};

  \node[ann] at (-0.55,-0.36) {$g_3$};
  \node[cell] at (0.95,-0.36) {$I$};
  \node[cell] at (1.70,-0.36) {$I$};
  \node[cell] at (2.45,-0.36) {$X$};
  \node[cell, anchor=west] at (3.05,-0.36) {$+$};

  \draw[semithick] (-0.55,-0.60) -- (4.10,-0.60);

  \node[ann] at (-0.55,-0.95) {$Z_a=g_1g_2g_3\in\mathcal{S}$: deterministic, $r_0=0$};
  \node[ann] at (-0.55,-1.38) {$Z$ on $d_1$ flips $g_2\to-g_2$, so $Z_a\to-Z_a$};

\end{tikzpicture}%
}
\caption{The stabilizer tableau of Figure~\ref{fig:stim-frames} immediately before
measurement.  The state is the joint $+1$ eigenstate of $g_1,g_2,g_3$, and the
measured observable $Z_a=g_1g_2g_3$ belongs to the stabilizer group.  Its outcome
is therefore fixed at $r_0=0$.  The relation
$g_1=+Z_aX_{d_1}X_{d_2}$ ties $Z_a$ to the data parity.  The frame produced by
the injected $Z$ error on $d_1$ flips $g_2$ and hence $Z_a$, matching the frame
propagation in Figure~\ref{fig:stim-frames}\,\circled{2}.}
\label{fig:tableau}
\end{figure}

\subsection{Quasiprobability Decomposition}
\label{subsec:qpd}

QPD extends stabilizer simulation to non-Clifford
processes.  A channel $\mathcal{E}$ is written as a real linear combination of
efficiently simulable stabilizer channels, $\mathcal{E}=\sum_j q_j\mathcal{S}_j$
with $q_j\in\mathbb{R}$, where a stabilizer channel may contain Clifford gates,
Pauli measurements, classical feed-forward, and Pauli
resets~\cite{Bennink_Ferragut_Humble_Laska_Nutaro_Pleszkoch_Pooser_2017}.
Trace preservation requires $\sum_j q_j=1$, and allowing some $q_j$ to be
negative is what makes the representation exact rather than an approximate
mixture.  The cost of that exactness is measured by the decomposition's one-norm
and its negativity,
\begin{equation}
  \gamma(\mathcal{E})=\sum_j|q_j|,
  \qquad
  \eta(\mathcal{E})=\sum_{q_j<0}|q_j|=\frac{\gamma-1}{2}.
  \label{eq:qpd-negativity}
\end{equation}

Sampling assembles these local decompositions into a global estimator.  Consider
an initial state $\rho$, channels $\mathcal{E}^{(1)},\ldots,\mathcal{E}^{(M)}$,
and an observable $O$, so that $F=\operatorname{Tr}[O\,\mathcal{E}^{(M)}\cdots\mathcal{E}^{(1)}(\rho)]$.  At noise site $k$ in the circuit, we sample branch $i_k$ with probability $p_{i_k}^{(k)}=|q_{i_k}^{(k)}|/\gamma_k$, where $\gamma_k=\sum_j|q_j^{(k)}|$.  The selected branches form a stabilizer circuit whose output we denote $f(\mathbf{i})$.  With $\Gamma=\prod_{k=1}^{M}\gamma_k$ and $w(\mathbf{i})=\Gamma\prod_{k=1}^{M}\operatorname{sgn}(q_{i_k}^{(k)})$, the Monte Carlo estimator 
\begin{equation*}
\widehat{F}_N=\frac{1}{N}\sum_{s=1}^{N}w(\mathbf{i}^{(s)})f(\mathbf{i}^{(s)})
\end{equation*}
is unbiased: $\mathbb{E}[\widehat{F}_N]=F$.  For a probability-valued observable, the variance obeys $\operatorname{Var}(\widehat{F}_N)\leq\Gamma^2/N$, so an additive-error estimate requires
\begin{equation}
  N=O\!\left(\frac{\Gamma^2}{\epsilon^2}
      \log\frac{1}{\delta}\right)
  \label{eq:qpd-samples}
\end{equation}
shots, where $\epsilon$ is the target additive precision and $\delta$ is the
statistical confidence parameter, i.e.,
$\Pr(|\widehat{F}_N-F|>\epsilon)\leq\delta$
\cite{Pashayan_Wallman_Bartlett_2015,Bennink_Ferragut_Humble_Laska_Nutaro_Pleszkoch_Pooser_2017}.
QPD is therefore exact in expectation, but its statistical cost is
multiplicative in the local one-norms: because $\Gamma$ is a product over sites,
a circuit-level noise model with one site per gate makes $\Gamma$ grow with
circuit size even when each $\gamma_k$ is close to $1$.

We use two examples of non-Pauli sources in this work: coherent over-rotation and thermal relaxation. Both decompose
into three terms, exactly
one of which is not a Pauli channel.  A coherent over-rotation about a Pauli axis $A$ gives
$
  \mathcal{R}_A(\Theta)=a\,\mathcal{I}+b\,\mathcal{A}+
  |\sin\Theta|\,\mathcal{S}_A,
  \quad
  \gamma=|\cos\Theta|+|\sin\Theta|,
$
where $\mathcal{S}_A$ is conjugation by $\sqrt{A}$ and $a,b$ are given by
Eq.~\eqref{eq:rotation-table}.  For $0<|\Theta|<\pi/2$ the negative weight sits
on $b$, so $\gamma>1$.  The thermal-relaxation decomposition is in
Appendix~\ref{sec:app-thermal-qpd}.  Both channels have non-Pauli branches,
which Stim cannot simulate directly.

\subsection{Pauli Twirling Approximation}

A common alternative is to approximate a general error channel by its Pauli
twirl.  For the $n$-qubit Pauli basis
$\mathcal{P}_n=\{I,X,Y,Z\}^{\otimes n}$, the twirl of $\Lambda$ is
\begin{equation*}
  \widetilde{\Lambda}(\rho)=4^{-n}\sum_{P\in\mathcal{P}_n}P^{\dagger}\Lambda(P\rho P^{\dagger})P.
\end{equation*}
If $\Lambda(\rho)=\sum_{P,Q}\chi_{PQ}P\rho Q^{\dagger}$, twirling removes the
off-diagonal process terms and leaves
$\widetilde{\Lambda}(\rho)=\sum_{P\in\mathcal{P}_n}\chi_{PP}P\rho P^{\dagger}$
with $\chi_{PP}\geq0$ and $\sum_{P}\chi_{PP}=1$.  The result is a positive Pauli
mixture, so $\gamma=1$ and the channel is compatible with direct Pauli-frame
sampling \cite{Geller_Zhou_2013}.

The throughput is therefore recovered by changing the channel rather than by
simulating it in a novel fashion.  For example, conjugation by
\begin{equation*}
  R_Z(\theta)=\cos(\theta/2)I-i\sin(\theta/2)Z
\end{equation*}
is replaced by
\begin{equation*}
  \widetilde{\mathcal{R}}_Z(\rho)=\cos^2(\theta/2)\,\rho+\sin^2(\theta/2)\,Z\rho Z,
\end{equation*}
which discards the coherent cross terms.  Using the twirled
channel as a surrogate for an untwirled device introduces a model bias, and the
twirl may additionally omit non-unital effects, leakage, loss, and coherent
accumulation. QPD instead preserves the original channel in expectation, at the
cost quantified by Eq.~\eqref{eq:qpd-samples}.

\paragraph{PTA vs. Coherent error example.}
Zhou et al.\ analyze coherent crosstalk in stabilizer measurements
\cite{Zhou_Ji_Ding_2025}, and their Eqs.~(5)--(6) make the mechanism explicit.
On a weight-four $X$ check, each data-basis component $x$ acquires a
state-dependent phase $\phi(x)=-\tfrac{1}{2}\sum_i\theta_i(-1)^{x_i}$.  The
final ancilla Hadamard then adds or subtracts the contributions of $x$ and
$\bar{x}=x\oplus(1,1,1,1)$, so for input amplitudes $\alpha_x$ and
$a_x=\alpha_xe^{i\phi(x)}$, the syndrome probability contains
\begin{equation}
  |a_x\pm a_{\bar{x}}|^2=|a_x|^2+|a_{\bar{x}}|^2
  \pm2\operatorname{Re}(a_xa_{\bar{x}}^*).
\end{equation}
Depending on the relative phase, the cross term either \textbf{enhances} or \textbf{suppresses}
the syndrome probability.  PTA replaces the coherent paths with a classical Pauli
mixture and so retains only the first two terms.  It therefore cannot reproduce
this interference even though
the twirled and coherent channels are matched on the single-location error rate.
\section{Motivation}
\label{sec:motivation}

\subsection{Realistic Noise Is a Systems Requirement}

Noisy-circuit simulation feeds three tasks in the fault-tolerant design loop.
Each task takes the noise model as an input, so the model's accuracy is part of
that task's result.
Decoder design uses simulated syndromes across codes, logical operations, and
circuit-noise models.
Recent neural and predecoder studies draw that data from algorithmic circuits,
surface codes, and color codes
\cite{BonillaAtaides_Gu_Yelin_Lukin_2025,Chamberland_Olle_Li_Thornton_Baratta_2026,Olle_Chamberland_Li_Baratta_2026}.
Experiments use simulation to set decoder priors, to identify logical failure
mechanisms, and to test whether measured error suppression matches a device
model.
That use is reported for below-threshold superconducting surface-code
experiments and for fault-tolerant neutral-atom demonstrations
\cite{GoogleQuantumAI_2025,Bluvstein_et_al_2026}.
Architecture studies use it to turn a hardware error budget into a code choice,
a compilation strategy, a decoder-throughput target, and a resource estimate
\cite{Tripier_et_al_2026,Cain_et_al_2026,Webster_et_al_2026,Campbell_et_al_2026}.

No single stochastic Pauli model fully captures the error mechanisms that real
hardware platforms exhibit. Superconducting transmons are subject to relaxation,
dephasing, coherent calibration errors, crosstalk, persistent leakage, and
occasional correlated events
\cite{Hashim_et_al_2021,Miao_et_al_2023,GoogleQuantumAI_2025}. Trapped-ion
error budgets include photon scattering, motional heating and dephasing, and
coherent laser-control errors
\cite{Ballance_Harty_Linke_Sepiol_Lucas_2016,LeBlond_Groszkowski_Lietz_Seck_Bennink_2025}.
Neutral-atom arrays combine coherent Rydberg-control errors with scattering,
leakage, and atom loss, and observed loss carries erasure information that a
Pauli-only model cannot represent
\cite{Wu_Kolkowitz_Puri_Thompson_2022,Bluvstein_et_al_2026}. Which mechanism
dominates is itself not fixed: it varies with gate type, idle duration,
calibration, and QEC protocol, so a model fitted at one operating point need
not hold at another.

Simplified models can consequently alter architectural conclusions. Exact and
phase-sensitive studies report differences between non-Clifford gates and their
Clifford proxies, between thermal relaxation and its Pauli approximation, and
between coherent and twirled detector error models
\cite{Chase_Labib_2026,Garner_Myers_Wang_Stein_Liu_Li_2025,Takou_Brown_2025}.
\emph{These results do not show that Pauli twirling is always inaccurate. They
show that its accuracy cannot be known in advance, which is the motivation for more accurate noise simulators such as
\quasistim{}.} 

\subsection{The Two QPD Bottlenecks}

QPD provides an unbiased estimator for general local qubit channels, at two
costs that compound for circuits of interest.

The first is the shot count, which scales as $\Gamma^2$, where
$\Gamma=\prod_i\gamma_i$ is the product of channel one-norms
(Eq.~\eqref{eq:qpd-samples}). A syndrome-extraction circuit contains many weak
non-Pauli sites, and $\Gamma$ accumulates multiplicatively over them, so it grows
with circuit size even when every $\gamma_i$ is close to one.

The second is the cost of executing a single shot. Stim amortizes its tableau
simulation because all shots share one noiseless reference; under QPD they do
not, since each sampled branch pattern could potentially select a different stabilizer circuit
and therefore a different reference. Conventional QPD consequently builds that
circuit and runs a tableau simulation on it once per shot. Reuse would require
two shots to draw the same pattern, but a pattern comprises an independent
branch choice at every site, so the number of distinct patterns grows
exponentially with the number of sites, and collisions are effectively absent.
Each shot thus pays for its own circuit construction and tableau reference,
precisely on the circuits for which $\Gamma^2$ already demands a large $N$.

\subsection{Design Goals and Scope}

\quasistim{} is designed to preserve the QPD estimator of
\S\ref{subsec:qpd}, including its branch probabilities and signed weights, and
to bring per-shot throughput close to Stim's Pauli-frame sampler.
The circuits it is built for are QEC circuits with many weak non-Pauli sites
and a large shot budget.
Extended-tableau and state-vector simulators address a different class of circuits:
those with a universal gate set under Pauli noise.
\section{Design}
\label{sec:methodology}

QPD costs a factor $\Gamma^2$ in shots and a tableau simulation per shot, and
\quasistim{} reduces each with a separate component. \quasifold{} is a compiler
frontend that rewrites noise sites so that $\Gamma$ falls, which lowers the
number of trajectories a target accuracy requires. \deltaframe{} is a sampling
backend that precomputes branch-dependent references once, so that each
trajectory costs a frame update rather than a tableau run. The two also
reinforce one another, since every site \quasifold{} removes shrinks the
pairwise sweep \deltaframe{} performs during preprocessing.

\newsavebox{\qfinputbox}
\sbox{\qfinputbox}{%
  \begin{quantikz}[font=\scriptsize, row sep={0.70cm,between origins},
    column sep=0.14cm, thin lines]
    \lstick{$q_0$}
      & \gate{H}
      & \gate[style={fill=qscream!85}]{R_Z(\theta_1)}
      & \ctrl{1}
      & \gate[style={fill=qsorange!45, dashed}]{\text{\texttt{Z\_ERR}}(p)}
      & \gate[style={fill=qscream!85}]{R_Z(\theta_2)}
      & \gate[style={fill=qsorange!45, dashed}]{D(p')}
      & \gate{M}
      & \\
    \lstick{$q_1$}
      &
      &
      & \targ{}
      &
      & \gate[style={fill=qscream!85}]{R_Z(\theta_3)}
      &
      & \gate{M}
      &
  \end{quantikz}}

\newsavebox{\qfoutputbox}
\sbox{\qfoutputbox}{%
  \begin{quantikz}[font=\scriptsize, row sep={0.70cm,between origins},
    column sep=0.18cm, thin lines]
    \lstick{$q_0$}
      & \gate{H}
      & \ctrl{1}
      & \gate[style={fill=qsred!45}]{\mathcal{C}(\Theta,p,p')}
      & \gate{M}
      & \\
    \lstick{$q_1$}
      &
      & \targ{}
      &
      & \gate{M}
      &
  \end{quantikz}}

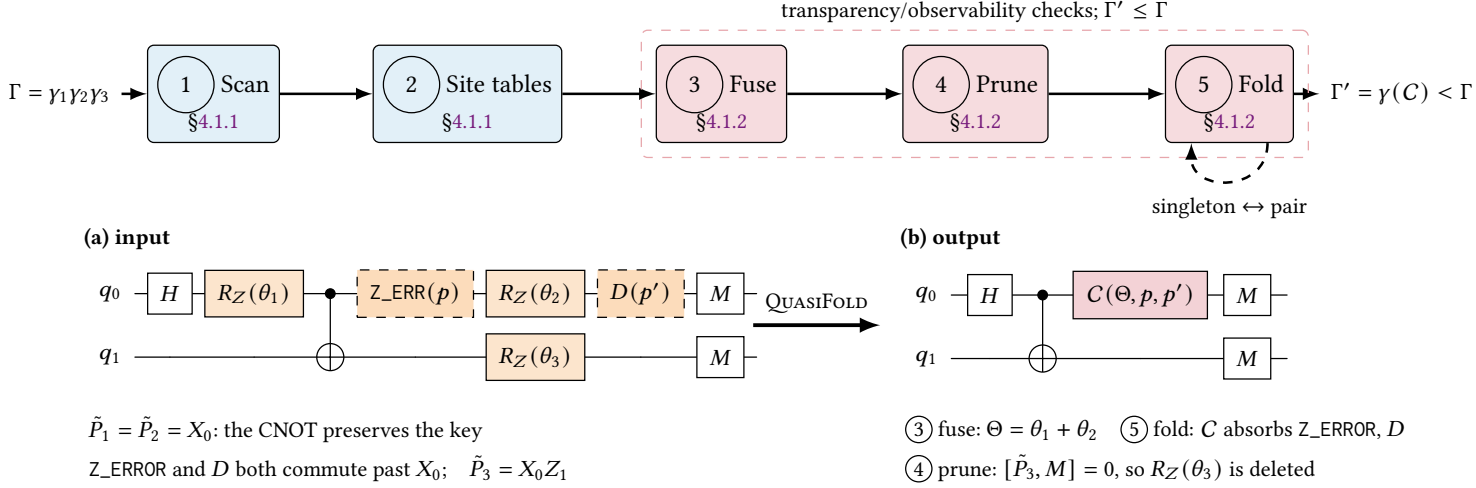
\begin{figure*}[t]
\centering
\scalebox{1.2}{%
\begin{tikzpicture}[
    font=\footnotesize,
    stage/.style={draw, rounded corners=2pt, minimum height=6.2mm,
                  inner xsep=3pt, align=center, fill=qsblue!35},
    pass/.style={stage, fill=qsred!35},
    flow/.style={-latex, thick},
    ann/.style={font=\scriptsize, anchor=west},
  ]

  \node[stage] (scan)  at (1.0,0)    {\circled{1} Scan\\[-1pt]\scriptsize \S\ref{sec:qf-repr}};
  \node[stage] (repr)  at (3.8,0)    {\circled{2} Site tables\\[-1pt]\scriptsize \S\ref{sec:qf-repr}};
  \node[pass]  (fuse)  at (6.6,0)    {\circled{3} Fuse\\[-1pt]\scriptsize \S\ref{sec:qf-fuse}};
  \node[pass]  (prune) at (9.4,0)    {\circled{4} Prune\\[-1pt]\scriptsize \S\ref{sec:qf-prune}};
  \node[pass]  (fold)  at (12.2,0)   {\circled{5} Fold\\[-1pt]\scriptsize \S\ref{sec:qf-fold}};

  \node[font=\scriptsize, anchor=east] (gin) at ($(scan.west)+(-0.28,0)$)
       {$\Gamma=\gamma_1\gamma_2\gamma_3$};
  \draw[flow] (gin) -- (scan);
  \node[font=\scriptsize, anchor=west] (gout) at ($(fold.east)+(0.28,0)$)
       {$\Gamma'=\gamma(\mathcal{C})<\Gamma$};
  \draw[flow] (fold) -- (gout);

  \draw[flow] (scan)  -- (repr);
  \draw[flow] (repr)  -- (fuse);
  \draw[flow] (fuse)  -- (prune);
  \draw[flow] (prune) -- (fold);

  \node[draw=qsred!90, dashed, rounded corners=2pt, fit=(fuse)(prune)(fold),
        inner sep=4.5pt] (passbox) {};
  \node[font=\scriptsize] at ($(passbox.north)+(0,0.20)$)
       {transparency/observability checks; $\Gamma'\leq\Gamma$};
  \draw[flow, dashed]
       ($(fold.south)+(0.42,0)$) .. controls +(0,-0.55) and +(0,-0.55) ..
       ($(fold.south)+(-0.42,0)$);
  \node[font=\scriptsize] at ($(fold.south)+(0,-0.72)$)
       {singleton $\leftrightarrow$ pair};

  \begin{scope}[yshift=-2.55cm]
    \node[ann, font=\scriptsize\bfseries] at (-0.55,0.95) {(a) input};
    \node[anchor=west] at (-0.50,0.00) {\usebox{\qfinputbox}};

    \node[ann] at (-0.5,-1.15) {$\tilde{P}_1=\tilde{P}_2=X_0$: the CNOT preserves the key};
    \node[ann] at (-0.5,-1.58) {\texttt{Z\_ERROR} and $D$ both commute past $X_0$;\quad $\tilde{P}_3=X_0Z_1$};
  \end{scope}

  \draw[flow, line width=1.2pt] (6.95,-2.55) --
        node[above, font=\scriptsize] {\textsc{QuasiFold}} (8.35,-2.55);

  \begin{scope}[xshift=9.0cm, yshift=-2.55cm]
    \node[ann, font=\scriptsize\bfseries] at (-0.55,0.95) {(b) output};
    \node[anchor=west] at (-0.50,0.00) {\usebox{\qfoutputbox}};

    \node[ann] at (-0.5,-1.15)
         {\circled{3} fuse: $\Theta=\theta_1+\theta_2$\quad
          \circled{5} fold: $\mathcal{C}$ absorbs \texttt{Z\_ERROR}, $D$};
    \node[ann] at (-0.5,-1.58)
         {\circled{4} prune: $[\tilde{P}_3,M]=0$, so $R_Z(\theta_3)$ is deleted};
  \end{scope}

\end{tikzpicture}%
}
\caption{\quasifold{} lowers the QPD $\Gamma$ by rewriting sites.  In
the worked example, the two $R_Z$ sites are separated by a CNOT, yet they carry
the same common-frame axis $X_0$, so stage \circled{3} fuses them by angle
arithmetic; the stochastic $Z$-error chunk \texttt{Z\_ERROR} and the
depolarizing chunk $D$ both commute with that axis, so stage \circled{5}
absorbs them into the surviving
site; and $R_Z(\theta_3)$ commutes with every measurement it can reach, so stage
\circled{4} deletes it.  Three sites and two noise chunks become one composite
site, and $\Gamma$ falls from
$\gamma_1\gamma_2\gamma_3$ to $\gamma(\mathcal{C})$.}
\Description{A five-stage pipeline scans and represents noise sites, fuses
same-axis rotations, prunes unobservable rotations, and folds Pauli noise.
A two-qubit example shows the corresponding site rewrites.}
\label{fig:quasifold-flow}
\end{figure*}

\subsection{\quasifold{} Frontend: Structure-Aware QPD Reduction}
\label{sec:quasifold}
\label{sec:qf-overview}

The shot count at a fixed accuracy target scales as
$\Gamma^2=(\prod_i\gamma_i)^2$, so any rewrite that reduces $\Gamma$ pays for
itself quadratically. Such rewrites exist because the one-norm is
submultiplicative under composition,
$\gamma(\mathcal{E}_2\circ\mathcal{E}_1)\leq\gamma(\mathcal{E}_1)\gamma(\mathcal{E}_2)$,
for the composed representation \cite{Kulikov_Yashin_Fedorov_Kiktenko_2024}.
Sampling two sites independently costs the product, whereas sampling one
composite drawn from their combined coefficients costs the left-hand side. Where
the inequality is strict, replacing the pair by the composite reduces cost
without changing the estimator's expectation.

\quasifold{} searches for that situation greedily, with three passes that fuse
rotations sharing an axis, delete rotations no measurement can observe, and
absorb Pauli noise into a neighboring site. Every pass must satisfy two
conditions: it must leave the quasiprobability-weighted measurement distribution
unchanged, and it must not raise $\Gamma$; a rewrite is accepted only when it
lowers $\Gamma$ or removes a site. Figure~\ref{fig:quasifold-flow} shows the
resulting pipeline, whose stage numbers this subsection uses throughout.
Stages \circled{1}--\circled{2} build a representation in which both conditions
can be checked, and stages \circled{3}--\circled{5} are the three passes.
Algorithm~\ref{alg:quasifold} in Appendix~\ref{sec:app-qf-search} gives the full
procedure.

\subsubsection{Representation and transparent paths (\circled{1}--\circled{2})}
\label{sec:qf-repr}
\label{sec:qf-transport}

Deciding whether two sites can be combined requires answering three questions:
whether they are compatible, whether the circuit between them permits the move,
and whether the result is cheaper. Stages \circled{1}--\circled{2} provide the
ability to assess all three.

Compatibility is settled by a change of frame. The scan tracks the inverse
Clifford prefix $C_{<i}$ and maps each site axis $P_i$ to
$\tilde{P}_i=C_{<i}^{\dagger}P_iC_{<i}$, so that two rotations separated by
Cliffords carry the same key exactly when they rotate about the same physical
axis. Their relative orientation in that frame fixes the sign each contributes
when they are fused.

Cost is settled by a common algebra. Rotations, damping channels, and Pauli
channels are all represented as signed coefficient tables listing stabilizer
branches and their weights. Composing two channels multiplies their tables and
adds the weights of any branches that coincide, and the one-norm is evaluated
only after that addition. This ordering is what makes a composite cheaper than
its factors: coinciding branches of opposite sign cancel before their magnitudes
are summed. Appendix~\ref{sec:app-qf-algebra} gives the multiplication rules.

The move itself is permitted only along a transparent path. Cliffords are
transparent by construction, since the common-frame key has already accounted
for them, while measurements, resets, composites, and blocking channels
terminate a composition block. Every other channel on the path needs a
pair-specific test. A Pauli channel is transparent when commuting it past the
rotation leaves the composed channel unchanged, and a rotation is transparent
when the two axes commute. Damping requires the stricter test of checking each
sampled branch rather than the channel as a whole, because two channels can
commute as maps while a pair of their branches does not.
Appendix~\ref{sec:app-qf-transport} gives the tests.

\subsubsection{Channel rewrites (\circled{3}--\circled{5})}
\label{sec:qf-rewrites}

\textbf{Coherent fusion (\circled{3}).}
\label{sec:qf-fuse}
Sites sharing a common-frame axis and joined by transparent paths form a group
$\mathcal{G}$, which fusion replaces with the single rotation
$\Theta_{\mathcal{G}}=\sum_{j\in\mathcal{G}}s_j\theta_j$, where
$s_j\in\{-1,+1\}$ is the site's orientation relative to the group axis. The
rewrite always reduces the site count, and it reduces $\Gamma$ whenever the
summed rotation carries a smaller one-norm than the product over the group,
which the pass verifies before accepting. Opposing angles are the favorable
case, since they partially cancel in the sum.

\textbf{Observability pruning (\circled{4}).}
\label{sec:qf-prune}
A rotation site can be deleted outright under either of two conditions: its
common-frame axis commutes with every measured observable reachable from it, or
an intervening reset removes its remaining support. In both cases the site
cannot affect any sampled outcome, so deleting it is exact and its one-norm
factor drops out of $\Gamma$ entirely. This can be the strongest of the three rewrites,
because it removes a factor rather than shrinking one.

\textbf{Noise folding (\circled{5}).}
\label{sec:qf-fold}
Folding absorbs a chunk of Pauli noise into a neighboring site, subject to three
conditions: both lie in the same measurement/reset block, the path between them
is transparent, and $\gamma_{\mathrm{new}}<\gamma_{\mathrm{old}}$. The third
condition is what makes the pass nontrivial. A Pauli channel has one-norm $1$,
so folding one is never useful in isolation; it becomes useful when the fold
introduces branches that cancel negative weights already present in the
neighboring non-Pauli table, which lowers the combined one-norm and in favorable
cases returns it to $1$. A noisy composite can therefore need strictly fewer
trajectories than the same channels sampled independently. The greedy search is
bounded, which keeps preprocessing tractable but can miss a better combination.
Appendices~\ref{sec:app-qf-algebra} and~\ref{sec:app-qf-search} detail the
coefficient updates and the candidate selection.

\begin{figure*}[t]
\centering
\scalebox{1.2}{%
\begin{tikzpicture}[
    font=\footnotesize,
    panel/.style={draw, dashed, rounded corners=2pt, inner sep=4pt},
    box/.style={draw, rounded corners=1.5pt, minimum height=5.0mm,
                inner xsep=3pt, font=\scriptsize, align=center},
    gnode/.style={draw, circle, minimum size=5.0mm, inner sep=0pt,
                  font=\scriptsize},
    cell/.style={draw, fill=white, minimum size=3.9mm, inner sep=0pt,
                 font=\scriptsize},
    head/.style={font=\scriptsize\bfseries, anchor=west},
    ann/.style={font=\scriptsize, anchor=west},
    cost/.style={font=\scriptsize\itshape, anchor=west},
    flow/.style={-latex, thick},
  ]

  \begin{scope}
    \node[head] at (0.0,1.45) {\circled{1} reference basis};
    \node[box, fill=qsblue!35] (r0) at (1.5,0.55) {one noiseless run $\Rightarrow$
                                   $r_0\in\mathbb{F}_2^{L}$};
    \node[box, fill=qsblue!35] (di) at (1.5,-0.30) {$\Delta_i(a)=r(i{=}a)\oplus r_0$};
    \node[font=\scriptsize] at (1.5,-0.92) {one per site, per digit};
    \node[panel, draw=qsblue!85, fit=(r0)(di)] (pa) {};
    \node[cost] at (0.0,-1.55) {$1+\sum_i k_i$ tableau references};
  \end{scope}

  \draw[flow] (3.35,0.10) -- (3.90,0.10);

  \begin{scope}[xshift=4.30cm]
    \node[head] at (-0.20,1.45)
         {\circled{2} pair sweep $\rightarrow$ \circled{3} block tables};

    \node[gnode, fill=qscream!85] (n1) at (0.45,0.50) {1};
    \node[gnode, fill=qscream!85] (n4) at (3.05,0.50) {4};
    \node[gnode, fill=qscream!85] (n8) at (5.65,0.50) {8};
    \node[gnode, fill=qspurple!70] (n2) at (1.35,0.50) {2};
    \node[gnode, fill=qspurple!70] (n3) at (2.15,0.50) {3};
    \draw[thick] (n2) -- (n3);
    \node[gnode, fill=qsgreen!70] (n5) at (3.95,0.50) {5};
    \node[gnode, fill=qsgreen!70] (n6) at (4.75,0.50) {6};
    \node[gnode, fill=qsgreen!70] (n7) at (4.35,-0.22) {7};
    \draw[thick] (n5) -- (n6);
    \draw[thick] (n6) -- (n7);
    \draw[thick] (n5) -- (n7);

    \node[draw=qspurple!90, dashed, rounded corners=2pt, fit=(n2)(n3), inner sep=2pt]
         (bA) {};
    \node[draw=qsgreen!90, dashed, rounded corners=2pt, fit=(n5)(n6)(n7), inner sep=2pt]
         (bB) {};
    \node[font=\scriptsize] at (1.75,1.16) {$\mathcal{B}_1$};
    \node[font=\scriptsize] at (4.35,1.16) {$\mathcal{B}_2$};

    \node[ann] at (-0.20,-0.92)
         {probed conflict: $\Delta_{ij}(a,b)\neq\Delta_i(a)\oplus\Delta_j(b)$};
    \node[panel, fit=(n1)(n8)(bA)(bB)] (pb) {};
    \node[cost] at (-0.20,-1.55)
         {$O(M^2)$ pair tests; $\textstyle\prod_{i\in\mathcal{B}}(1+k_i)-1$
          joint refs per block};
  \end{scope}

  \draw[flow] (10.30,0.10) -- (10.85,0.10);

  \begin{scope}[xshift=11.00cm]
    \node[head] at (-0.05,1.45) {\circled{4} sample time};

    \node[font=\scriptsize, anchor=east] at (0.02,0.92) {lanes};
    \foreach \r/\y in {1/0.50, 2/0.10, 3/-0.30} {
      \node[font=\scriptsize, anchor=east] at (0.02,\y) {\r};
    }
    \foreach \x in {0.25,0.65,1.05,1.45,1.85,2.25,2.65,3.05} {
      \node[cell] at (\x,0.50) {0};
      \node[cell] at (\x,0.10) {0};
      \node[cell] at (\x,-0.30) {0};
    }
    \node[cell, fill=qspurple!70] at (0.65,0.50) {1};
    \node[cell, fill=qsgreen!70] at (1.85,0.10) {1};
    \node[cell, fill=qsgreen!70] at (2.65,0.10) {2};
    \node[cell, fill=qscream!85] at (0.25,-0.30) {1};
    \node[cell, fill=qspurple!70] at (1.05,-0.30) {1};
    \node[cell, fill=qscream!85] at (3.05,-0.30) {1};
    \node[font=\scriptsize] at (1.65,0.92) {site $1\ldots M$};

    \node[box, fill=qsred!35, anchor=west] at (-0.05,-0.95)
         {$\widehat{r}(p)=r_0\oplus\bigoplus_{\mathcal{B}}
           \Delta_{\mathcal{B}}(p_{\mathcal{B}})$};
    \node[cost] at (-0.05,-1.55)
         {lookup and XOR; no tableau here};
  \end{scope}

\end{tikzpicture}%
}
\caption{\deltaframe{} reference precomputation replaces one tableau simulation per shot
with one tableau simulation per \emph{interaction}, and sampling then assembles
a reference by lookup and XOR.  \circled{1} A single noiseless run fixes $r_0$,
and one run per site digit fixes the single-site deltas.  \circled{2} Each site
pair is tested for linearity; a violation becomes an edge.  \circled{3} Each
connected component of the resulting graph is a block whose joint
configurations are enumerated exactly, so the exponential cost is paid per
cluster rather than over the global pattern space $|\Omega|=\prod_i(1+k_i)$.
\circled{4} At sample time, lanes carrying different patterns share one batch:
lane 2 fires two sites of the same block and consumes one joint entry, while
lane 3 fires three sites in three different blocks and XORs one entry from
each.  All sample operations are $O(1)$ in the Pauli frame.}
\label{fig:deltaframe}
\end{figure*}
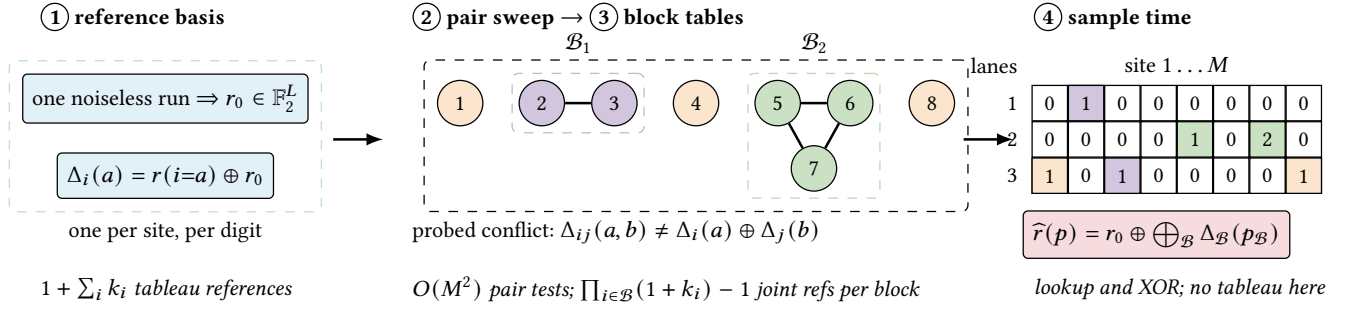

\subsection{\deltaframe{} Backend: Reference Precomputation}
\label{sec:deltaframe-compile}

Let the rewritten circuit contain $M$ non-Pauli sites, with site $i$
contributing $k_i$ non-Pauli branches. Each trajectory is then identified by a
pattern $p=(d_1,\ldots,d_M)$, $d_i\in\{0,\ldots,k_i\}$, in which digit $0$
selects the site's pure Pauli branch and a nonzero digit selects one of its
stabilizer branches.

Distinct patterns select distinct stabilizer circuits, and this is what breaks
Stim's split. Under stochastic Pauli noise every branch is a Pauli, so all shots
execute the same Clifford circuit and share the single noiseless reference from
which each recovers its record as $\mathrm{rec}_s=r_0\oplus f_s$. A QPD branch
inserts a Clifford or a reset, so the correct reference becomes a function of
$p$. Recomputing it per shot would reinstate exactly the tableau cost the split
was built to avoid.

The way out is that references compose additively when branches do not
interact. Two sites whose branches act independently perturb the record
independently, so their reference offsets combine by XOR; this holds
unconditionally between Pauli branches. Non-Pauli branches may interact, and
then the joint offset is not the XOR of the individual offsets and no
$\mathbb{F}_2$ combination of them predicts it. Interactions of that kind are
nevertheless local, because they originate in local gates and are cut by the
intervening measurement and reset layers. \deltaframe{} exploits this by
precomputing single-site offsets, detecting which sites interact, and
enumerating jointly only inside each interacting group. The alternative of
caching complete patterns is exponential in $M$.

Figure~\ref{fig:deltaframe} shows the pipeline that follows: compute
single-site offsets, test site pairs and group the interacting ones into
blocks via union-find, enumerate each block's configurations, and assemble every lane's
reference at sample time.

\subsubsection{Reference basis (\circled{1})}
\label{sec:df-basis}

Take $r_0$ to be the reference of the all-zero pattern. Activating digit $a$ at
site $i$ alone then defines the offset $\Delta_i(a)=r(i{=}a)\oplus r_0$.
Building the basis costs one baseline run plus one run per nonzero site digit,
and each offset is stored as a sparse set of measurement-bit differences.

\subsubsection{Conflict blocks (\circled{2}--\circled{3})}
\label{sec:df-blocks}

Two sites are joined by an edge when their joint reference offset differs from
the XOR of their single-site offsets, which is the signature of an interaction.
The connected components of the graph are the blocks. Enumerating the branch
configurations within each block $\mathcal{B}$ then permits any pattern's
reference to be predicted as
\begin{equation}
  \widehat{r}(p)=r_0\oplus
  \bigoplus_{\mathcal{B}}\Delta_{\mathcal{B}}(p_{\mathcal{B}}).
  \label{eq:block-reference}
\end{equation}
Each block entry records that block's full local branch configuration, not
merely its pairwise contributions, so interactions inside a block are captured
exactly.

The cost of this enumeration is exponential in block size rather than in $M$,
which helps only if the components stay small: a sparse conflict graph with one
large component is no cheaper than caching patterns outright. Our target
circuits fall in the favorable regime, because local gates and the measurement
and reset layers confine interactions. The largest block across every evaluation
circuit is five sites, and thermal relaxation produces no conflicts at all,
leaving each site its own block. Table~\ref{tab:block-structure} in
Appendix~\ref{sec:app-deltaframe} gives the measured counts and size
distributions.

\subsection{\deltaframe{} Backend: Batched Sampling}
\label{sec:deltaframe-sampling}

With the basis and block tables in place, sampling proceeds much as Stim's SIMD
Pauli-frame sampling does, with one difference: a single batch may hold lanes
carrying different QPD patterns. The shared circuit is traversed once per batch.
Per-lane branch masks select which conditional Pauli and non-Pauli updates
apply, block lookups supply each lane's predicted reference, and XORing that
reference with the lane's accumulated measurement flips yields its sampled
record.

For a fixed circuit and batch width, sampling work is therefore linear in the
shot count. What the design does not do is eliminate tableau work; it relocates
it into preprocessing, where it appears as a quadratic site-pair sweep and an
enumeration exponential in block size. Appendix~\ref{sec:app-deltaframe} gives
the cost model and the enumeration limits.

One consequence of relocating that work is that
Eq.~\eqref{eq:block-reference} is not universally exact. Interactions above
pairwise order, and interactions between blocks, can invalidate a predicted
reference, and the pair sweep detects neither.
Section~\ref{sec:eval-accuracy} therefore compares predicted references against
direct tableau records and reports both the mismatch rate and the resulting bias
in the logical error rate.
\section{Evaluation}
\label{sec:evaluation}

The evaluation proceeds as follows.  \S\ref{sec:eval-setup} gives the
evaluation setup.  \S\ref{sec:eval-speedup} measures the two speedup components
separately: how much \quasifold{} lowers the shot count, and how much faster
\deltaframe{} executes one trajectory; the overall speedup is the product of the
two.  \S\ref{sec:eval-deltaframe-overhead} reports the one-time cost of
\deltaframe{} reference precomputation.  \S\ref{sec:eval-gap} studies where Pauli twirling and exact simulation
(the latter enabled by \quasistim{}) disagree, across memory, logical
operations, and magic-state cultivation.  That is where the gap matters most
for long-term FTQC studies.  It also includes a rerun of a published
thermal-relaxation study.  \S\ref{sec:eval-accuracy} measures the
accuracy of \deltaframe{} (note that \quasifold{} is exact by construction).

\subsection{Experimental Setup}
\label{sec:eval-setup}

\subsubsection{Circuits.}
We evaluate on three classes of circuits, chosen to cover the regimes a
fault-tolerant design loop has to validate: quantum memory, which isolates
error suppression as distance grows; logical operations, which add lattice
surgery and basis transitions to the set of error paths; and magic-state
cultivation, which is post-selected and therefore sensitive to error paths that
a memory experiment averages over.  Memory circuits for surface, color, and
bivariate-bicycle codes are generated with the qLDPC library
\cite{Perlin_qLDPC}.  Logical-operation circuits come from FTPrimitiveBench
\cite{Kan_Harkness_Du_Rofougaran_Garner_Liu_Mao_Stein_2026}.  Magic-state
cultivation circuits follow the protocol of Gidney et al.~\cite{Gidney_Shutty_Jones_2024}.
Table~\ref{tab:benchmark-circuits} summarizes the evaluated circuits' properties.

\begin{table}[thbp]
  \centering
  \scriptsize
  \setlength{\tabcolsep}{2.8pt} 
  \caption{Properties of the evaluation circuits.}
  \label{tab:benchmark-circuits}
  \begin{tabular}{@{}llrrrrr@{}}
    \toprule
    Circuit & Config. & Qubits & Depth & Ops & Pauli Sites & Non-Pauli Sites \\
    \midrule
    SC mem. & $d=3$, each 2q         & 17  & 19   & 138    & 218    & 72  \\
    SC mem. & $d=5$, defects, 2q     & 49  & 31   & 690    & 974    & 65  \\
    SC mem. & $d=7$, defects, 2q     & 97  & 43   & 1{,}946 & 2{,}610 & 133 \\
    \midrule
    Color mem. & $d=3$, idle         & 10  & 48   & 122    & 383    & 45  \\
    Color mem. & $d=5$, idle         & 28  & 80   & 638    & 1{,}711 & 199 \\
    Color mem. & $d=7$, idle         & 55  & 112  & 1{,}838 & 4{,}611 & 536 \\
    \midrule
    BB mem. & $d=6$, each 2q         & 144 & 67   & 3{,}600 & 6{,}552 & 288 \\
    \midrule
    Logical $H$ & $d=3$, each 2q     & 17  & 21   & 147    & 114    & 72  \\
    Logical $S$ & $d=3$, each 2q     & 44  & 79   & 1{,}116 & 871    & 138 \\
    MSC e2e    & $d_1=3$, $d_2=15$   & 454 & 148  & 14{,}552 & 50{,}763 & 81 \\
    \bottomrule
  \end{tabular}
\end{table}

\subsubsection{Noise Models.}
We use two families of non-Pauli noise.  For coherent noise, we follow two
models: that of~\cite{Zhou_Ji_Ding_2025}, in which every two-qubit gate is
followed by an \texttt{RZZ} crosstalk rotation, and that
of~\cite{LeBlond_Groszkowski_Lietz_Seck_Bennink_2025}, in which every idling
qubit receives an \texttt{RZ} coherent error before each measurement round.  For
incoherent non-Pauli noise we use thermal relaxation, parameterized by the
duration $\tau$ for which a qubit idles before measurement relative to its
relaxation time, so that a sweep of $\tau/T_1$ moves the channel from nearly
trivial to strongly damping; unless stated otherwise we take $T_1=T_2$.  The two
families differ in the branch their decomposition draws: a Clifford rotation
for coherent noise and a reset for damping.  This difference matters for cost
(\S\ref{sec:eval-cost-model}) as well as for accuracy.  On top of either family,
we add uniform background depolarizing noise using the qLDPC
library~\cite{Perlin_qLDPC}.

Because $\Gamma$ is a product over channel sites, instrumenting every site of a
high-distance circuit exceeds the compute available to us, so some
high-distance benchmarks use a defect-qubit model; this is an artifact of our
resource budget rather than a limitation of the method.  Unlike the unrestricted
model, which applies coherent noise after every two-qubit gate, the
defect-qubit model applies \texttt{RZZ} rotations only when a selected
defective data qubit participates in the gate.  Pauli twirling a $Z$ or $ZZ$
rotation of angle $\theta$ yields the corresponding $Z$ or $ZZ$ error with
probability $\sin^2(\theta/2)$.  Table~\ref{tab:rz-twirl} lists that
probability for the coherent angles used below.

\begin{table}[t]
  \centering
  \scriptsize
  \setlength{\tabcolsep}{2pt}
  \caption{Pauli-twirl error probability $p_Z=\sin^2(\theta/2)$ of each
  coherent $R_Z$ angle.}
  \label{tab:rz-twirl}
  \begin{tabular}{@{}lcccccc@{}}
    \toprule
    $\theta$ & $0.002\pi$ & $0.005\pi$ & $0.006\pi$ & $0.01\pi$ & $0.014\pi$ & $0.02\pi$ \\
    \midrule
    $p_Z$ & $9.87\times10^{-6}$ & $6.17\times10^{-5}$ & $8.88\times10^{-5}$ & $2.47\times10^{-4}$ & $4.84\times10^{-4}$ & $9.87\times10^{-4}$ \\
    \bottomrule
  \end{tabular}
\end{table}

\subsubsection{Simulation Infrastructure.}
All measurements are executed on Perlmutter CPU nodes with two AMD EPYC 7763
processors (AVX2), 128 cores, and 512\,GB of memory.

\subsubsection{Baselines.}
We compare against three arms.  The sampling baseline is the standard
\mcsampler{} construction for stabilizer-channel
QPD~\cite{Bennink_Ferragut_Humble_Laska_Nutaro_Pleszkoch_Pooser_2017}: each
trajectory is an independent tableau simulation, with no reference precomputation.  At every
QPD site a random branch is sampled, the corresponding stabilizer circuit is
built, and Stim's tableau simulator executes it via \texttt{do\_circuit}; the
process repeats in full for each shot.  This arm is exact in expectation, so it
isolates the cost of \deltaframe{} at matched accuracy.  The twirled baseline
is the PTA arm: each non-Pauli channel is replaced by its Pauli twirl and the
resulting circuit is sampled with Stim's Pauli-frame sampler.  This is the
arm practitioners actually use, and it is the reference for every gap we report
in \S\ref{sec:eval-gap}.  Finally, where the circuit is small enough, we check
both against Clifft, an exact non-stabilizer reference whose memory footprint
limits it to low distance; \S\ref{sec:eval-gap-memory} reports where that check
is available.

\subsection{Speedup}
\label{sec:eval-speedup}

The two components reduce different costs, so we measure them against different
quantities: \quasifold{} against the shot count $\Gamma^2$, and \deltaframe{}
against wall-clock time per trajectory.

\subsubsection{\quasifold{}: one-norm reduction}
\label{sec:eval-onenorm}

Because the shot count required for a fixed accuracy scales as $\Gamma^2$
(Eq.~\eqref{eq:qpd-samples}), the figure of merit for \quasifold{} is the ratio
$\Gamma_{\mathrm{before}}^2/\Gamma_{\mathrm{after}}^2$ against the same circuit
compiled with \quasifold{} disabled.  We measure it on three code families:
surface and color memory at $d=3,5,7,9$ and bivariate-bicycle memory at
$d=6,10$, in each case with $\mathrm{rounds}=d$.  Both noise models use
$R_Z(0.006\pi)$ and depolarizing noise at $p=10^{-3}$; the first places a
rotation at every two-qubit gate, and the second places one on each idle
qubit before measurement.

\begin{figure}[t]
  \centering
  \includegraphics[width=\columnwidth]{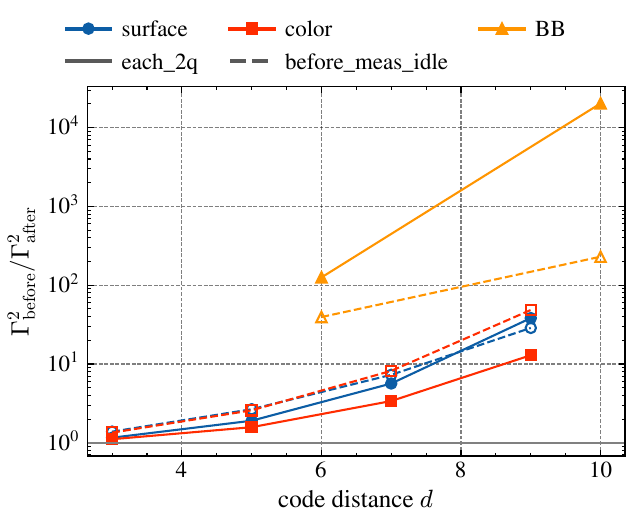}
  \caption{Shot-count reduction from \quasifold{}, as
  $\Gamma_{\mathrm{before}}^2/\Gamma_{\mathrm{after}}^2$ against the same
  circuit compiled without optimization.}
  \label{fig:quasifold}
\end{figure}

Figure~\ref{fig:quasifold} reports the measured ratio.  The reduction grows
with distance in every family, and the reason is structural: $\Gamma$ is a
product over channel sites, so each site the pass merges or removes contributes
a multiplicative saving, and the number of such sites grows with the circuit.
Across families at fixed distance, the ordering follows how much of the circuit
carries channel sites at all.  Bivariate-bicycle memory reduces the most
overall, because its higher encoding rate involves more qubits, and therefore
more instrumented gates, at a given distance than surface or color memory.

Which of the two noise models benefits more depends on the code, its schedule,
and the distance.  On surface memory, the two are close, and they cross:
rotations on idle qubits before measurement lead through $d=7$, and rotations
before every two-qubit gate overtake at $d=9$.  Two effects run against each
other.  A surface code idles boundary qubits at every measurement round, so the
idle placement has sites to work on at every distance, but those qubits are a
shrinking fraction of the patch as $d$ grows, whereas the number of two-qubit
gates grows with the patch area; the idle placement's site density therefore
falls with distance while the gate placement's does not.

\subsubsection{\deltaframe{}: cost per trajectory}
\label{sec:eval-cost-model}

\begin{figure}[h]
  \centering
  \includegraphics[width=.5\textwidth]{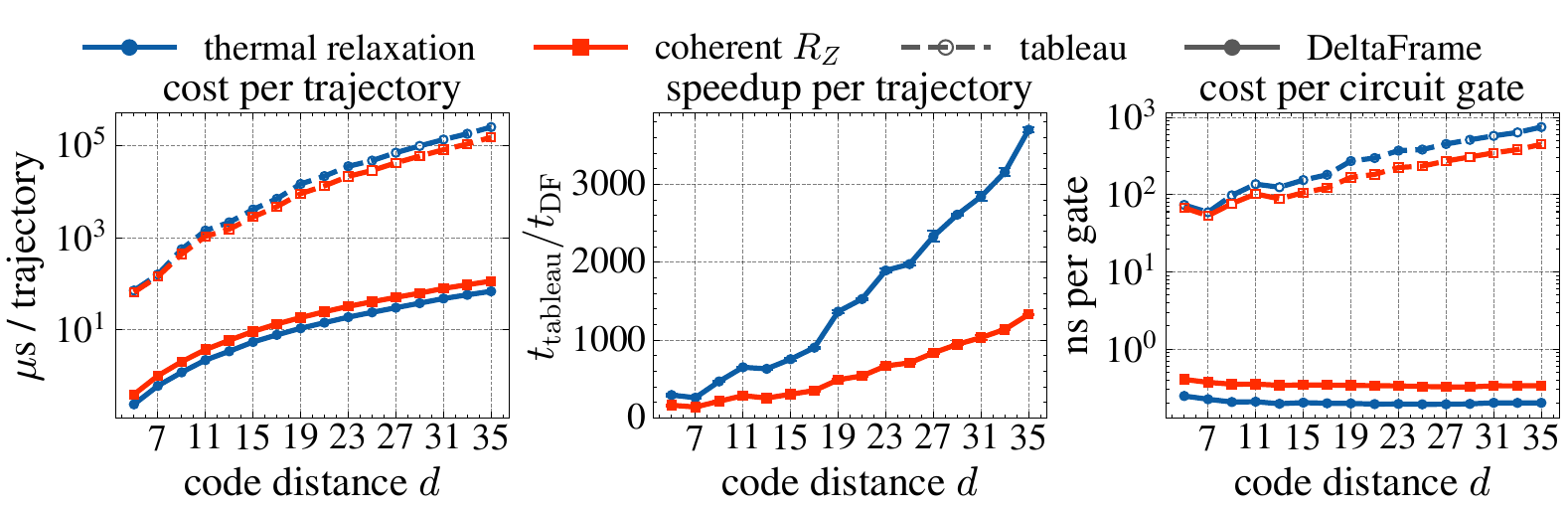}
  \caption{Cost per trajectory of \deltaframe{} and the \mcsampler{} tableau
  baseline on rotated surface $X$-memory, their ratio, and the same costs per
  gate.}
  \label{fig:singlecore}
\end{figure}

Figure~\ref{fig:singlecore} compares \deltaframe{} against the \mcsampler{}
tableau baseline on surface memory from $d=5$ to $d=35$, under coherent $R_Z$ on
every two-qubit gate and under thermal relaxation on qubits idling before
measurement.  At every distance, \deltaframe{} is two to three orders of
magnitude faster, because sampling operates on Pauli frames instead of tableaus.  The
gap widens with circuit size, from $166\times$ to $1331\times$ under $R_Z$
and from $297\times$ to $3702\times$ under damping, because \deltaframe{} holds
a constant cost per gate while the baseline grows with qubit count.  Concretely,
a tableau update touches $O(n)$ rows for a Clifford, $O(n^2)$ rows for a random
measurement, and $O(n)$ rows for a deterministic measurement on $n$ qubits. Since $n$
grows with the patch area, the baseline's per-gate cost rises with distance.  A frame update touches a fixed
number of bit-sliced words per gate regardless of $n$, so \deltaframe{}'s
per-gate cost is flat in distance, and the ratio between the two grows.

The same argument explains why the two noise families do not give the same
speedup, and here the branch each decomposition draws
(\S\ref{subsec:qpd}) is what matters.  Damping's non-Pauli branch is a reset,
which drives the tableau through a measurement and its $O(n^2)$ row reduction for a random measurement,
whereas the $R_Z$ branch is the single-qubit Clifford $\sqrt{Z}$ at $O(n)$.  The
baseline therefore pays substantially more per damping site than per coherent
site, while \deltaframe{} applies either branch as an $O(1)$ frame update and is
indifferent to which was drawn.

\subsection{Discrepancy between Twirling and Exact Simulation}
\label{sec:eval-gap}

With the speedup of \quasistim{}, we are able to systematically study the discrepancy between twirling and exact simulation.  We quantify that difference as the gap in
logical error rates between a PTA arm and a \quasistim{} arm on a set of QEC
circuits.  The disagreement is not a simple function of non-Pauli noise
strength: it depends on the circuit, the decoder, and the noise model, and the
experiments below vary each in turn.

\subsubsection{Memory}
\label{sec:eval-gap-memory}

We begin with memory circuits for the three code families of
\S\ref{sec:eval-setup}, sweeping the strength of one non-Pauli channel at a time
at fixed distance and fixed incoherent background noise.

\begin{figure}[t]
  \centering
  \includegraphics[width=.5\textwidth]{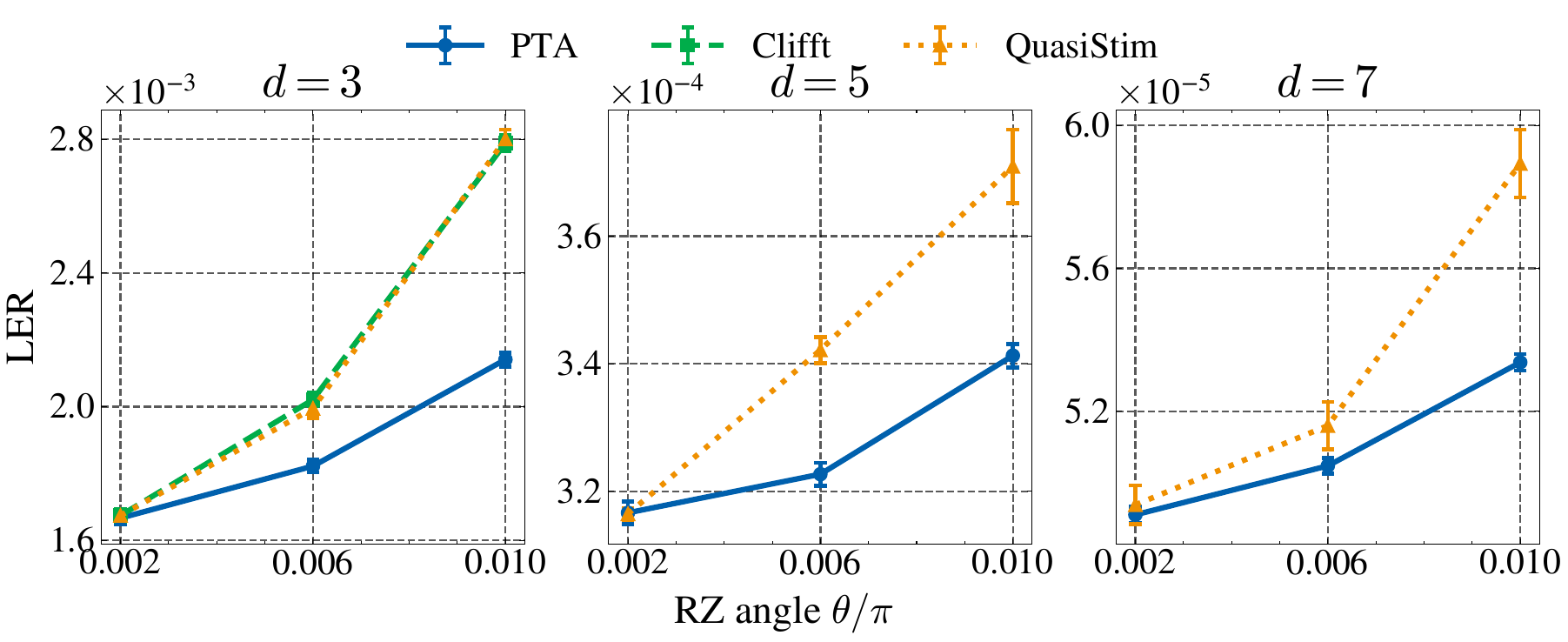}
  \caption{LER of rotated surface memory from \quasistim{} and PTA against the $R_Z$ angle.}
  \label{fig:surface-d357-rz}
\end{figure}

\paragraph{Coherent crosstalk.}
Figure~\ref{fig:surface-d357-rz} compares the logical error rate from
\quasistim{} and PTA on rotated surface memory, decoded in both arms with
PyMatching~\cite{Higgott2025sparseblossom}.  PTA systematically underestimates
the LER, and on a fixed circuit, the underestimate grows with coherent-noise
strength.  Two separate resource limits shape this experiment.  First, an
independent exact check against Clifft is available only at $d=3$, since Clifft
exhausts memory beyond that distance; the $d=5$ and $d=7$ points therefore rest
on \quasistim{} alone.  Second, \quasistim{} itself cannot run the unrestricted
coherent model at $d=5,7$ within our compute budget, because $\Gamma$ is a
product over $R_Z$ sites and the unrestricted model instruments every two-qubit
gate.  Those two distances consequently use the defect-qubit model, which
instruments far fewer sites. The measured gap should therefore be read as a lower bound: we
expect a larger gap under the unrestricted model.

\begin{figure}[t]
  \centering
  \includegraphics[width=.5\textwidth]{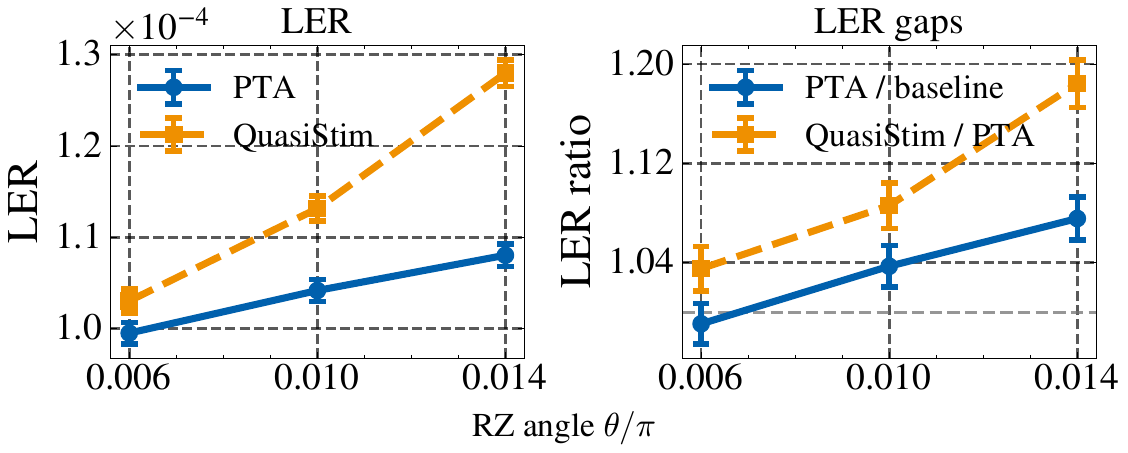}
  \caption{LER ratios on bivariate-bicycle memory against the $R_Z$
  angle: PTA over the $\theta=0$ baseline, and \quasistim{} over PTA.}
  \label{fig:bb-rz}
\end{figure}

We repeat the comparison on bivariate-bicycle memory in
Figure~\ref{fig:bb-rz}, again sweeping the rotation angle and measuring LER,
with both arms decoded by Relay-BP~\cite{mullerImprovedBeliefPropagation2025}
at \texttt{stop\_nconv}=2 and $\gamma_0=0.1$.  The figure plots two ratios of
logical error rates: PTA over the noiseless-rotation ($\theta=0$) baseline,
which shows how much LER the twirled model attributes to the coherent channel,
and \quasistim{} over PTA, which shows the excess the twirled model misses.  Both ratios
grow with the angle, and the second diverges from the first as the angle grows.
This is because coherent crosstalk can accumulate constructively at the amplitude
level across repeated QEC operations, causing superlinear LER growth, while PTA
removes this phase coherence so errors accumulate only probabilistically and the
LER grows roughly linearly. 

\begin{figure}[t]
  \centering
  \includegraphics[width=\columnwidth]{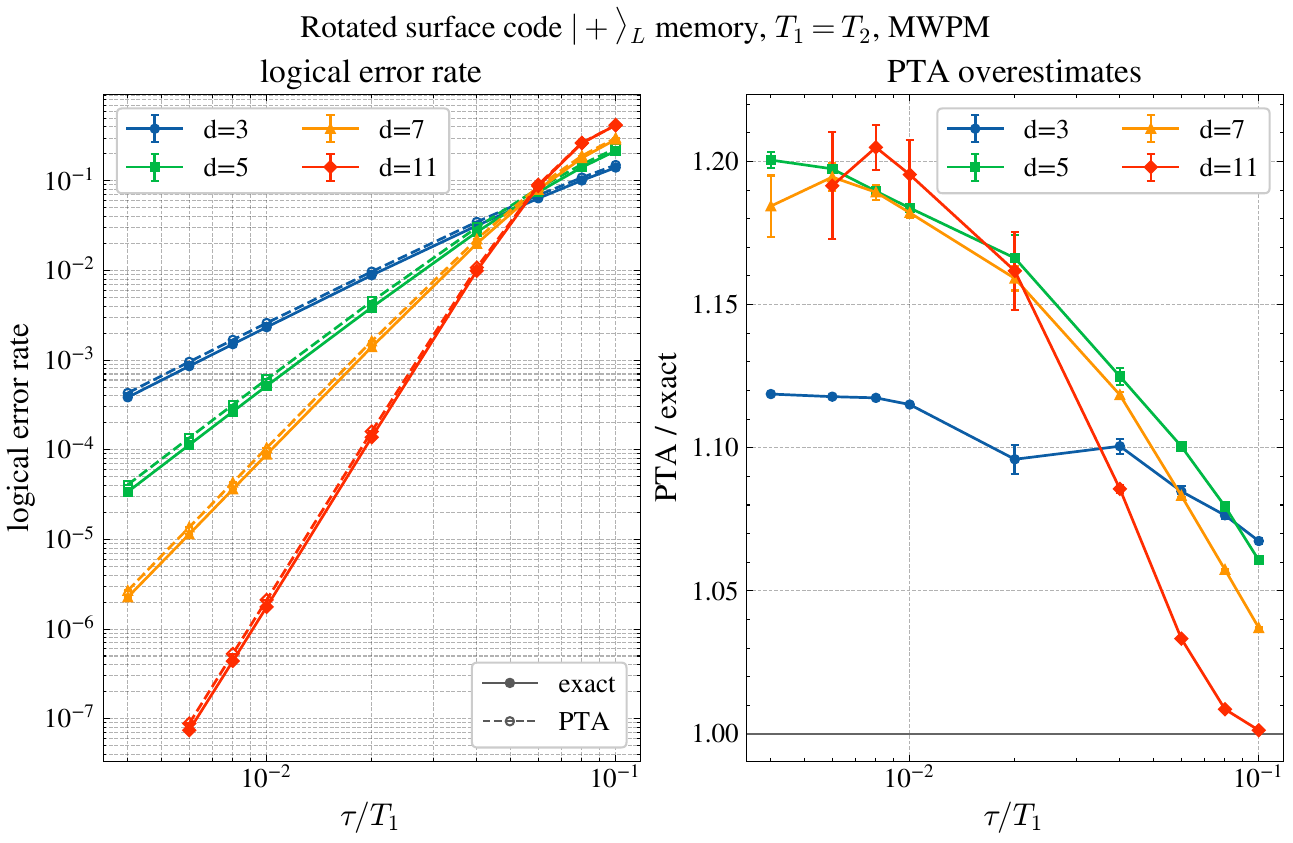}
  \caption{LER of rotated surface $\ket{+}_L$ memory from \quasistim{} and
  PTA, and their ratio, against $\tau/T_1$ under thermal relaxation alone.}
  \label{fig:damp-ler-surface}
\end{figure}

\paragraph{Thermal relaxation.}
We now turn to the non-Pauli incoherent noise.  We sweep the measurement-idle
duration $\tau/T_1$ on rotated surface-code $\ket{+}_L$ memory at $T_1=T_2$ with
$\mathrm{rounds}=d$ and no depolarizing noise background, so that thermal relaxation is
the only error source.
Figure~\ref{fig:damp-ler-surface} reports the LER of both arms and their ratio.
In contrast to the previous coherent crosstalk experiment, the gap grows as the non-Pauli noise weakens.

This is also the setting studied by Garner et al., who model thermal relaxation
for QEC and report that PTA distorts it
\cite{Garner_Myers_Wang_Stein_Liu_Li_2025}, so the sweep doubles as a
cross-check.  Our exact LER agrees with theirs, and our decompositions agree on
cost as well: theirs is positive whenever $T_2\leq T_1$, the same regime in which
ours requires no additional shots (\S\ref{sec:background}).  The
exact-versus-PTA gap we measure is much smaller than the one they report.  Since the
two exact estimates agree, the discrepancy likely stems from their
configuration of the twirled arm rather than from the exact estimation itself.\footnote{We have informed the authors of \cite{Garner_Myers_Wang_Stein_Liu_Li_2025} about this discrepancy.}  We obtain the exact result at an order of magnitude lower runtime.
The same sweep with a $p_{\mathrm{sto}}=5\times10^{-4}$ depolarizing
background is in Appendix~\ref{sec:app-damp-background}.

\begin{figure}[t]
  \centering
  \includegraphics[width=.5\textwidth]{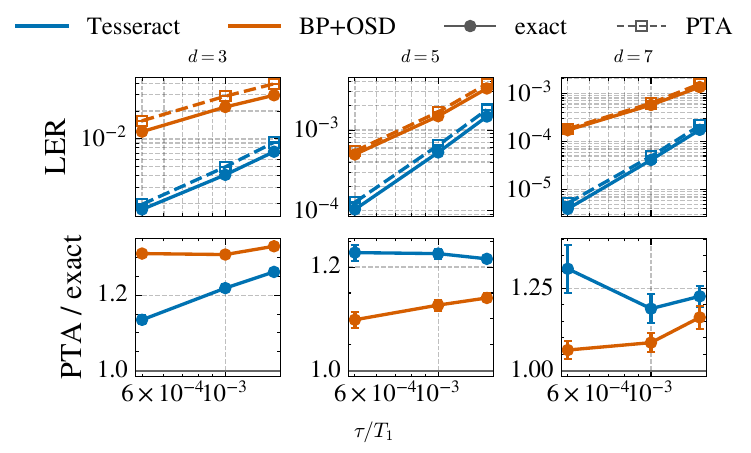}
  \caption{LER of color memory from \quasistim{} and PTA against $\tau/T_1$,
  decoded by Tesseract and by BP+OSD.}
  \label{fig:damp-ler-color}
\end{figure}

\paragraph{Dependence on the decoder.}
We further test whether the gap depends on the decoder by comparing two
decoders on color memory.  As shown in
Figure~\ref{fig:damp-ler-color}, Tesseract outperforms BP+OSD in every
setting we ran, by roughly an order of magnitude in LER, and it is also the more
sensitive of the two to thermal relaxation.  On both decoders PTA
\emph{overestimates} the LER, the opposite of most other experiments in this
section.  PTA therefore does not always err in one direction, and the sign as
well as the size of the gap depends on the decoder paired with it.

\subsubsection{Logical operations}
\label{sec:eval-gap-logical}

As fault-tolerant systems advance toward executing computation on encoded states, memory benchmarks alone become insufficient.  We therefore
measure the gap on two primitives from FTPrimitiveBench
\cite{Kan_Harkness_Du_Rofougaran_Garner_Liu_Mao_Stein_2026}: the transversal
Hadamard and the logical $S$ gate.  We lower the incoherent background to
$3\times10^{-4}$ because logical operations are more error-prone than memory, and we want to remain in the below-threshold regime.  The transversal Hadamard is close to
surface-code memory and is reported in Appendix~\ref{sec:app-transversal-h}.
We focus here on the $S$ gate.

\begin{figure}[t]
  \centering
  \includegraphics[width=\columnwidth]{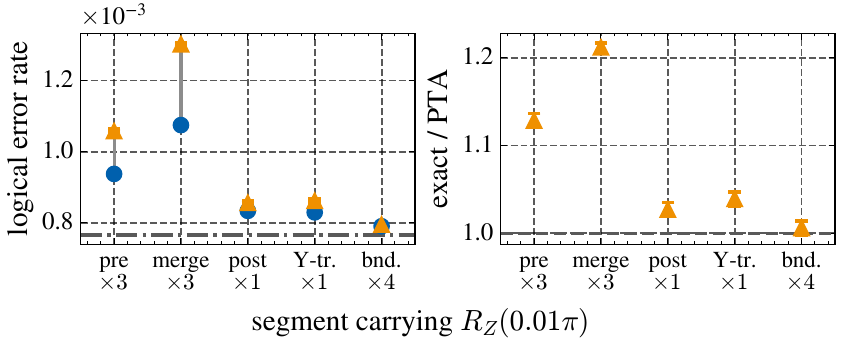}
  \caption{LER of the logical $S$ gate from \quasistim{} and PTA, with
  $R_Z(0.01\pi)$ confined to one segment at a time.}
  \label{fig:ftpb-sgate-segments}
\end{figure}

\paragraph{Logical $S$ gate.}
The $S$ gate is the least symmetric circuit we measure: lattice surgery around
one Y-basis transition.  The rounds around that transition form five segments.
Pre-rounds stabilize the patches before they join, merge rounds measure the
joint parity of the merged patch, post-rounds finish the split, a single
$Y$-transition round performs the basis change, and boundary rounds follow the
transition. Figure~\ref{fig:ftpb-sgate-segments} gives the number of rounds in
each segment.
Exact $R_Z(0.01\pi)$ on every site at once is intractable, because $\Gamma$
multiplies over sites.  We keep that rotation exact on one segment, Pauli-twirl
the crosstalk on the rest, and still simulate the full circuit. For example, the first
point keeps the three pre-rounds exact and twirls all other rounds.  Each segment is measured separately,
and the gap of the fully exact circuit is roughly the product of these per-segment gaps.
Figure~\ref{fig:ftpb-sgate-segments} shows that the error caused by Pauli twirling is not spread
evenly over the circuit.  The merge segment produces the largest gap, followed
by the pre-rounds, while on the boundary rounds exact and PTA almost agree.  The
merge is also the segment in which the joint parity measurement couples the two
patches, which is a plausible route for coherent error paths to become
correlated across the seam.

\subsubsection{Magic-state cultivation}
\label{sec:eval-gap-msc}
In FTQC, robust magic states are a key ingredient for universality.
Magic-state
cultivation (MSC)~\cite{Gidney_Shutty_Jones_2024} is a leading protocol for
preparing them.  MSC runs in three stages: injection prepares an initial
$\ket{T}$ state, cultivation raises the fault distance by cross-checking and
discarding any trial in which a detector fires, and escape grows the patch to
the target distance.  Because the middle stage discards rather than corrects,
MSC is the one benchmark here whose cost is set by a post-selection rate, which
makes it sensitive to error paths that memory experiments average away.  We
inject coherent crosstalk on every two-qubit gate in the injection and
cultivation stages, and measure the gap in LER per kept shot and its
dependence on the post-selection rate.

\begin{figure}[t]
  \centering
  \includegraphics[width=.5\textwidth]{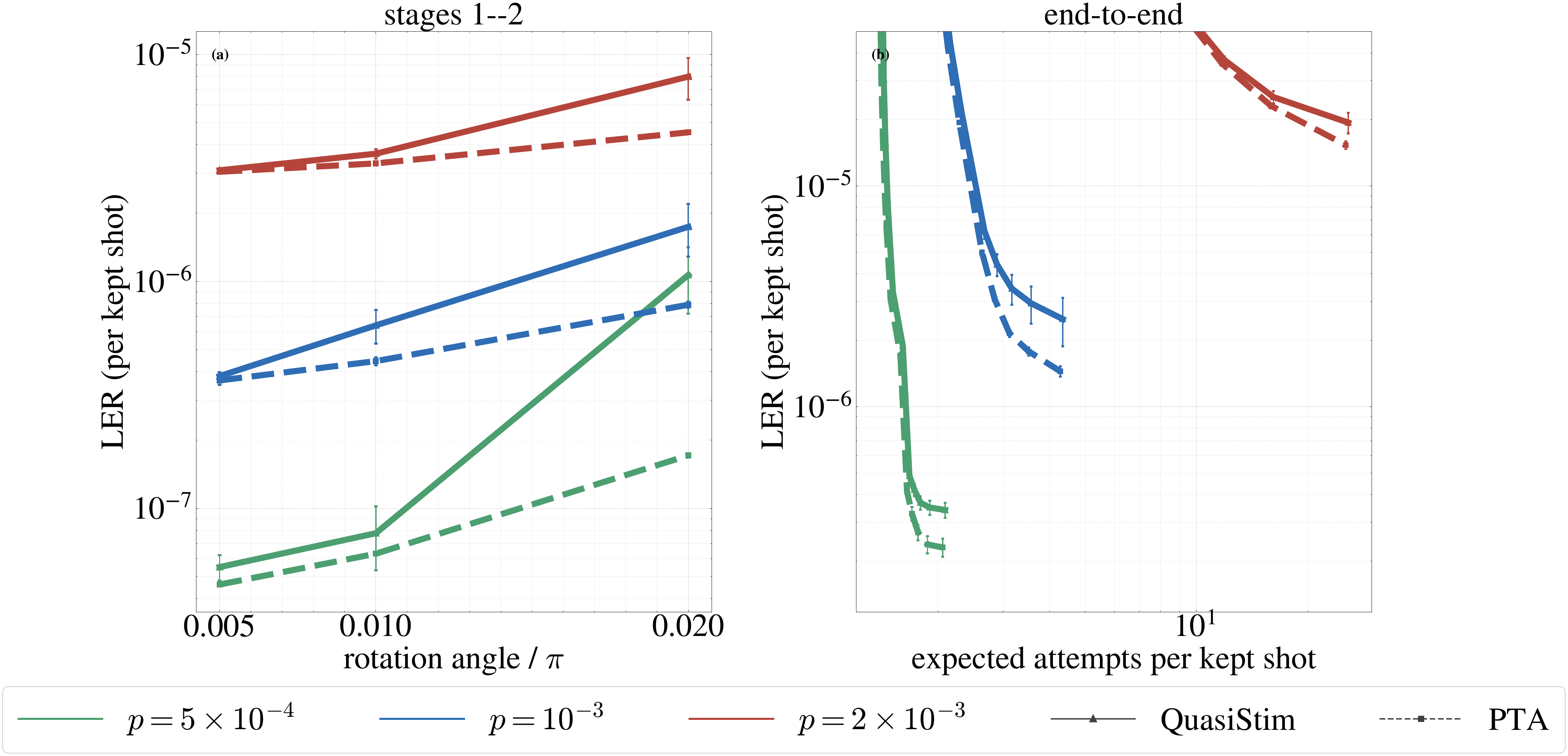}
  \caption{LER of magic-state cultivation from \quasistim{} and PTA, exact as
  solid triangles and PTA as dashed squares: stages 1 and 2 against the
  $R_Z$ angle in the left panel, and the end-to-end circuit against
  expected attempts per kept shot in the right panel.}
  \label{fig:msc-combined}
\end{figure}

Figure~\ref{fig:msc-combined} shows the LER gap between PTA and exact simulation
for stages 1 and 2 (left panel) and for the end-to-end circuit
(right panel).  The gap is most pronounced at low background noise, about $6\times$ in the
lowest-noise setting.  The end-to-end results in the right
panel show that the gap increases as the expected number of attempts per
successful shot rises, reflecting the increased impact of correlated errors
under more selective post-selection at lower target logical error rates.

\subsubsection{Predecoder case study}
A case study of the Pinball predecoder under coherent noise is in
Appendix~\ref{sec:app-predecoding}.

\subsection{\deltaframe{} Precomputation Overhead}
\label{sec:eval-deltaframe-overhead}

\deltaframe{} pays a one-time cost to build $r_0$, the single-site offsets and
the block tables, before any shot is drawn.  Table~\ref{tab:block-structure}
reports that cost on $128$ workers for every evaluation circuit.

Across all the benchmarks, the overhead is small and is dominated by the pair sweep.
Against sampling runs of millions of shots and above, this is negligible. The overhead grows with the number of sites, as well as the number and size of conflict blocks, as Eq.~\eqref{eq:deltaframe-compile-cost} in
Appendix~\ref{sec:app-deltaframe-cost} shows.

\subsection{Accuracy of \deltaframe{}}
\label{sec:eval-accuracy}

\deltaframe{} predicts $\widehat{r}(p)$ from precomputed block offsets rather
than simulating that pattern's tableau
(\S\ref{sec:deltaframe-compile}).  We measure the prediction on the evaluation
circuits.

For each circuit, we draw patterns from the sampling distribution and
compare $\widehat{r}(p)$ with a direct tableau simulation of the same pattern.
A draw is a \emph{mispredict} when the two records disagree.
Table~\ref{tab:deltaframe-accuracy} reports the resulting rates when sampling each circuit $200$ million times.

\begin{table}[htbp]
  \centering
  \small
  \caption{\deltaframe{} accuracy. Coherent noise is $R_Z(0.006\pi)$; damping is
  $\tau/T_1=10^{-2}$.}
  \label{tab:deltaframe-accuracy}
  \begin{tabular}{@{}llrr@{}}
    \toprule
    Circuit & Config. & Mispredict rate & Silent-error rate \\
    \midrule
    SC mem. & $d=3$, 2q & $1.1\times10^{-5}$ & $1.0\times10^{-6}$ \\
    SC mem. & $d=5$, 2q & $8.0\times10^{-6}$ & 0 \\
    SC mem. & $d=7$, 2q & $2.0\times10^{-5}$ & 0 \\
    BB mem. & $d=6$, 2q & 0 & 0 \\
    MSC 1+2 & $d_1=3$ & $3.9\times10^{-6}$ & 0 \\
    MSC 1+2 & $d_1=5$ & $1.7\times10^{-5}$ & 0 \\
    Logical $H$ & $d=3$, 2q & $2.5\times10^{-8}$ & $2.5\times10^{-8}$ \\
    Logical $H$ & $d=5$, 2q & $9.5\times10^{-7}$ & $3.3\times10^{-9}$ \\
    Logical $S$ & $d=3$, 2q & $6.2\times10^{-6}$ & $4.3\times10^{-6}$ \\
    Logical $S$ & $d=5$, 2q & $5.5\times10^{-6}$ & $2.7\times10^{-8}$ \\
    \midrule
    SC mem. & $d=3\text{--}9$, idle & 0 & 0 \\
    BB mem. & $d=6$, idle & 0 & 0 \\
    \bottomrule
  \end{tabular}
\end{table}

In a QEC circuit, a reference misprediction can affect the decoded LER whenever the predicted and
direct-tableau records produce different decoded logical-failure indicators.
Silent errors are defined as mispredictions that cause the observable to flip without any detector firing. Table~\ref{tab:deltaframe-accuracy} shows that the misprediction rates across all benchmarks are orders of magnitude lower than the logical error rate. Furthermore, the silent error rate is much lower than the misprediction rate and decreases with code distance; thus, \deltaframe{}'s precomputation approximation does not affect the logical error rate evaluation. Notably, the precomputation accuracy cost is exactly zero for all thermal-relaxation benchmarks.

\section{Related Work and Discussion}
\label{sec:discussion}

\subsection{Simulators}
Simulators differ in their hard scaling parameter: Hilbert-space dimension,
entanglement, magic, Pauli support, or QPD one-norm.  \quasistim{} occupies the
last of these, for large Clifford QEC circuits whose non-Pauli sites are many
and locally weak.
Tableau systems remain the Clifford baseline
\cite{Aaronson_Gottesman_2004,Gidney_2021}.  STABSim and cuQuantum accelerate
those kernels on GPU
\cite{Garner_Liu_Wang_Stein_Li_2025,Bayraktar_cuquantum_2023}.  \quasistim{}
could use these accelerations for \deltaframe{} reference precomputation and
sampling. Piveteau et al.\ and related work develop methods for reducing QPD
one-norms in error-mitigation and probability-estimation settings
\cite{Piveteau_Sutter_Woerner_2022,Kulikov_Yashin_Fedorov_Kiktenko_2024,Koukoulekidis_Kwon_Jee_Jennings_Kim_2022}.
Clifft and Tsim extend exact simulation to a small number of $T$ gates, at a
cost exponential in that non-Clifford resource
\cite{Chase_Labib_2026,Haenel_Luo_Zhao_2026}.  They spend the non-Clifford
budget on the circuit, not on the noise.  \quasistim{} targets the opposite
regime: many weak non-Pauli sites in a large Clifford circuit.  When the
coherent resource stays globally small, those methods should outperform local
QPD.
Fail-fast and dynamical subset sampling reduce the trajectories needed to
observe logical failure
\cite{Beverland_Carroll_Cross_Yoder_2025,Heussen_Winter_Rispler_Muller_2024}.
They change which trajectories are drawn to trigger more logical failure and are
complementary to \quasistim{}.  Prefix-sharing, sparse observable methods, and
other near-Clifford representations are in Appendix~\ref{sec:app-related}.

\subsection{Limitations and future work.}
\label{sec:limitations}
\deltaframe{} predicts most references, but blockwise and higher-order
interactions still produce mispredicts (\S\ref{sec:eval-accuracy}).
\quasifold{}'s search is greedy and can miss jointly beneficial groups.
Even after folding, $\Gamma^2$ is intractable for high non-Clifford volume.
Rare-event sampling, global negativity reduction, and hybridizing with Clifft
are natural next steps; we leave them to future work.  Leakage and qubit loss
remain out of scope; a mixed Pauli representation of leakage
\cite{ziad2025local} could be integrated at the sampling stage.

\begin{figure}[t]
  \centering
  \includegraphics[width=\columnwidth]{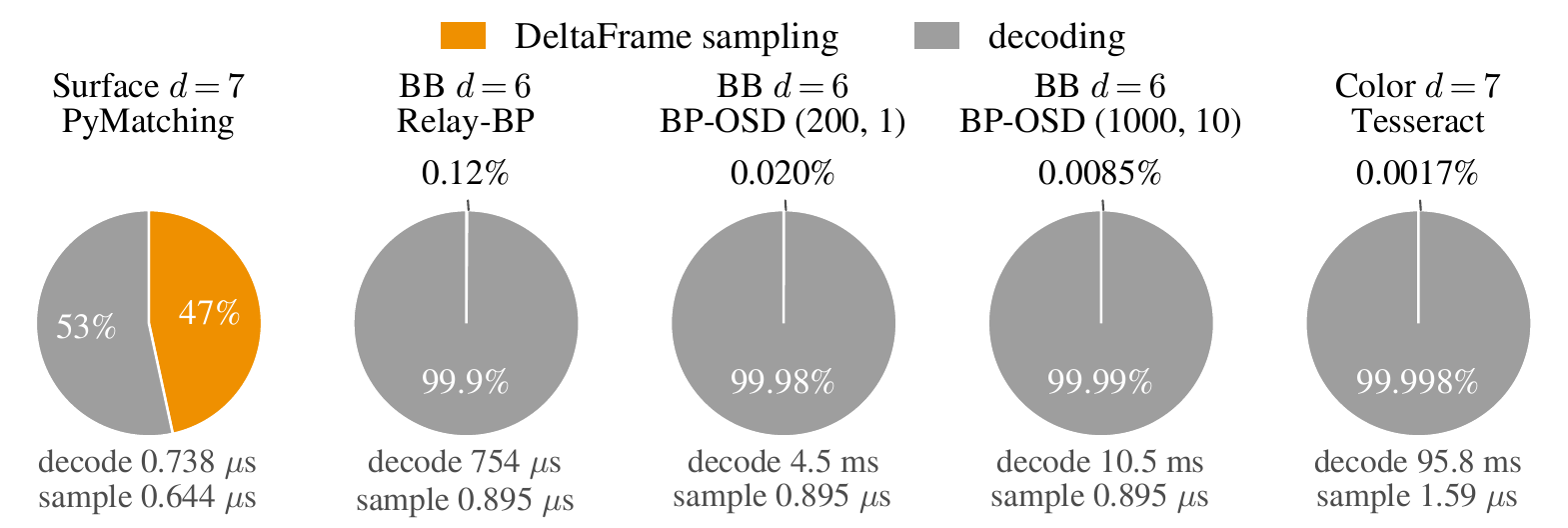}
  \caption{Time cost distribution per shot of \deltaframe{} sampling and of decoding the
  same shots, at $p_{\mathrm{sto}}=5\times10^{-4}$ and $R_Z(0.006\pi)$.
  BP-OSD is labeled (BP iterations, OSD
  order).}
  \label{fig:sample-vs-decode}
\end{figure}

A practical bottleneck for circuit-level QEC evaluation—and an amplified burden for \quasistim{} due to its $O(\Gamma^2)$ shot-count overhead—is decoding latency. As shown in Figure~\ref{fig:sample-vs-decode}, \deltaframe{} lowers sampling latency so substantially ($0.64$--$1.59,\mu$s per shot) that classical decoding overwhelmingly dominates single-shot execution time. While fast graph matching (e.g., PyMatching on surface codes) exhibits runtime comparable to sampling, decoders such as Relay-BP, BP-OSD, and Tesseract consume $99.8\%$ to over $99.99\%$ of the per-shot budget, requiring up to tens of milliseconds per syndrome. This asymmetry is not unique to \quasistim{}; it equally bounds standard Pauli simulation with Stim. Nonetheless, because exact QPD scales the required shot volume, developing high-throughput, accelerated decoders is critical for both classical QEC exploration and real-time physical control.

\section{Conclusion}\label{sec:conclusion}

We presented \quasistim{}, a scalable simulation framework for general error
channels beyond stochastic Pauli noise. Systematic benchmarks of fault-tolerant
circuits characterize the limitations of the Pauli twirling approximation and
the differences that emerge under more realistic hardware noise. These results
support QEC studies that evaluate architectures against the specified device
channel rather than its twirl.

\section*{Acknowledgements}
This material is based upon work supported by the U.S. Department of Energy, Office of Science, Office of Advanced Scientific Computing Research, Accelerated Research in Quantum Computing under Award Number DE-SC0025633. 
This research was, in part, supported by the DARPA HARQ Program, under Agreement No. HR0011-26-9-E096. 
This research used resources of the National Energy Research Scientific Computing Center, a DOE Office of Science User Facility supported by the Office of Science of the U.S. Department of Energy under Contract No. DE-AC02-05CH11231 using NERSC award NERSC DDR-ERCAP0035341. This research was, in part, funded by the U.S. Government. The views and conclusions contained in this document are those of the authors and should not be interpreted as representing the official policies, either expressed or implied, of the U.S. Government.
During the preparation of this work, the authors used Claude and ChatGPT for assistance with manuscript preparation. The authors reviewed and verified all results and writing, and take full responsibility for the accuracy and integrity of the final manuscript.

\bibliographystyle{ACM-Reference-Format}
\bibliography{real_sim_refs}

\appendix
\newpage
\section{Thermal-Relaxation Quasiprobability Decomposition}
\label{sec:app-thermal-qpd}

Thermal relaxation over a window $t$ with times $T_1$ and $T_2$ is the second
three-term source used in this work.  Write $s=e^{-t/2T_1}$ and
$d=e^{-t/T_\phi}$, where $T_\phi$ is the pure-dephasing time defined by
$1/T_\phi=1/T_2-1/2T_1$.  The channel decomposes as
\begin{equation}
  \Lambda_{T_1T_2}=c_0\mathcal{I}+c_1\mathcal{Z}+c_2\mathcal{R},
  \label{eq:thermal-qpd}
\end{equation}
where $\mathcal{R}$ is a reset to $\ket{0}$ and
$c_1=\tfrac{s}{2}(s-d)$ is the only negative coefficient.  Applying
Eq.~\eqref{eq:qpd-negativity} gives $\gamma=1-2c_1=1+s(d-s)$.  The one-norm
exceeds $1$ only when $T_2>T_1$, and a single channel's one-norm never exceeds
$1.25$.
For $T_2\leq T_1$ the weights are an ordinary probability distribution, so the
decomposition incurs no extra sampling cost.

\section{Thermal Relaxation with an Incoherent Background}
\label{sec:app-damp-background}

Damping alone is not a realistic error budget, so we extend the model of Garner
et al.~\cite{Garner_Myers_Wang_Stein_Liu_Li_2025} with a uniform depolarizing
background at $p_{\mathrm{sto}}=5\times10^{-4}$.  Thermal relaxation is applied
per time slice rather than per round, and the sweep widens to
$\tau/T_1\in[10^{-3},\,3\times10^{-1}]$, matching the no-background surface-code
sweep of \S\ref{sec:eval-gap-memory} except for those two changes and the added background.

\begin{figure}[thbp]
  \centering
  \includegraphics[width=\columnwidth]{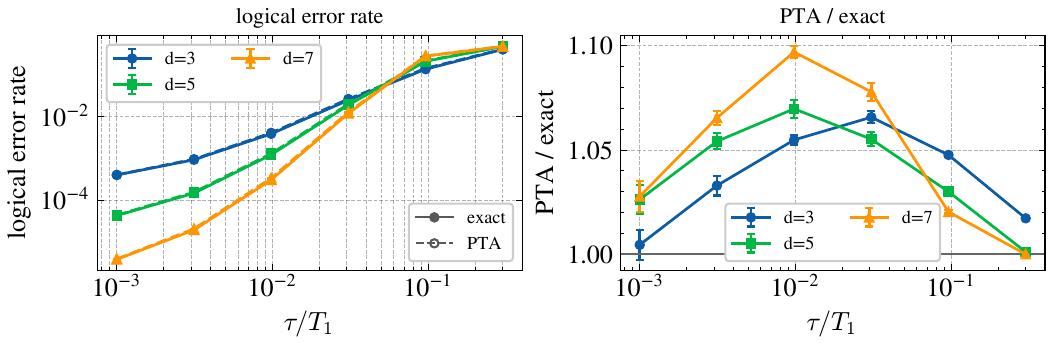}
  \caption{LER of rotated surface $\ket{+}_L$ memory from \quasistim{} and
  PTA, and their ratio, against $\tau/T_1$ with a
  $p_{\mathrm{sto}}=5\times10^{-4}$ depolarizing background.}
  \label{fig:damp-ler-surface-wide}
\end{figure}

Figure~\ref{fig:damp-ler-surface-wide} shows two changes relative to
Figure~\ref{fig:damp-ler-surface}.  The gap is suppressed: it peaks at
$1.066\times$ ($d=3$), $1.070\times$ ($d=5$), and $1.097\times$ ($d=7$),
against $1.12$--$1.21\times$ without the background.  It also stops being
monotonic in $\tau/T_1$, rising from $1.005$--$1.028\times$ at
$\tau/T_1=10^{-3}$ to that peak near $\tau/T_1=10^{-2}$, then falling to
$1.000$--$1.017\times$ at $\tau/T_1=0.3$.  We leave the theoretical analysis of
these results to future work.  These results provide further evidence that PTA cannot always
substitute for the exact channel.

\section{Transversal Hadamard}
\label{sec:app-transversal-h}

\begin{figure}[thbp]
  \centering
  \includegraphics[width=\columnwidth]{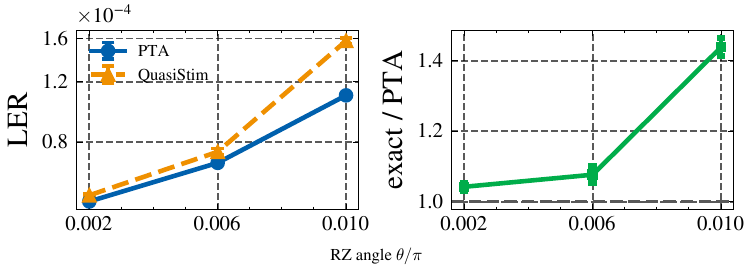}
  \caption{LER of the transversal Hadamard from \quasistim{} and PTA, against
  the $R_Z$ angle.}
  \label{fig:ftpb-transversal-h}
\end{figure}

On the transversal Hadamard from FTPrimitiveBench
\cite{Kan_Harkness_Du_Rofougaran_Garner_Liu_Mao_Stein_2026}, we apply the
two-qubit-gate crosstalk model and decode with PyMatching.  The incoherent
background is $3\times10^{-4}$, matching \S\ref{sec:eval-gap-logical}.
Figure~\ref{fig:ftpb-transversal-h} reports the LER of both arms against the
$R_Z$ angle.  The behavior matches surface-code memory: PTA systematically
underestimates the LER, and the gap grows with coherent-noise strength.  That
similarity is expected, since the transversal gate adds only single-qubit
operations on top of otherwise memory-like syndrome extraction, so nearly all
instrumented sites, and therefore nearly all error paths, are ones a memory
experiment already exercises.

\section{Predecoder Case Study}
\label{sec:app-predecoding}

Here we rerun a published predecoder evaluation under exact simulation
and report whether its findings change.
Pinball is a cryogenic predecoder evaluated under circuit-level Pauli noise
\cite{Knapen_Tao_Mack_Bruno_Saligane_Sylvester_Zhang_Ravi_2025}.  Its
lightweight rules correct non-complex errors on the spot and forward only the
syndrome data marked complex to the full decoder, which reduces the bandwidth
between the control processor and the quantum processor.

We use Stim's rotated surface-code $X$-memory at $d=3,5,7$ with $d$ rounds and
SI1000 background noise at $p=10^{-3}$.  We insert $R_Z(\theta)$ on CNOT
targets before the gate, sweeping $\theta/\pi\in\{0.002,0.006,0.01\}$.  At
$d=3$ every CNOT is instrumented; at $d=5,7$ only gates touching $d$ defect
data qubits on the logical observable's support are instrumented, so the columns
are not a matched distance-scaling experiment.  All arms use the same Pinball
rules.  We report two metrics: L1 coverage, the fraction of shots whose syndrome
Pinball classifies as non-complex, and L1 LER, the logical-failure probability
conditioned on L1 acceptance.  The first sets the bandwidth the predecoder saves;
the second sets whether that saving is sound.

\begin{figure}[t]
  \centering
  \includegraphics[width=\columnwidth]{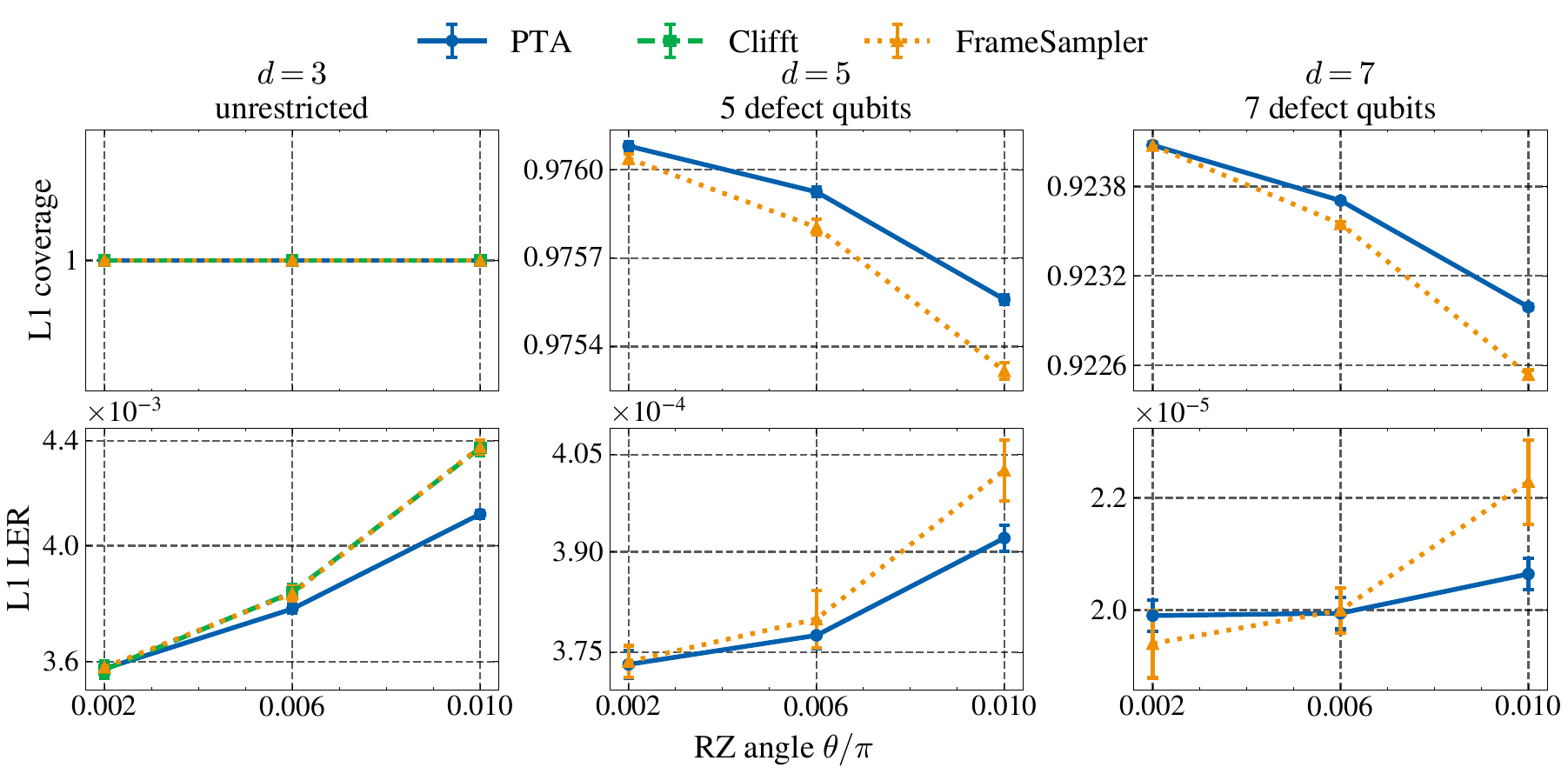}
  \caption{L1 coverage (\emph{top}) and L1 LER (\emph{bottom}) of Pinball on
  rotated surface $X$-memory at $d=3,5,7$, from \quasistim{} and PTA, against
  the injected $R_Z$ angle.}
  \label{fig:pinball-predecoder}
\end{figure}

Figure~\ref{fig:pinball-predecoder} separates coverage from correctness.  At
$d=3$ the patch is small enough that Pinball's rules absorb every syndrome, so
coverage is $1$ under all three arms.  At $d=5$ and $d=7$ the arms disagree:
PTA is the optimistic one, and the overstatement grows with the angle.  We read
this as coherent error paths producing correlated syndromes that the rules
classify as complex, which twirling breaks into the isolated defects they
absorb.  PTA is also optimistic on L1 LER: at $\theta=0.01\pi$ the exact L1 LER
exceeds that of PTA at both $d=5$ and $d=7$, and at smaller angles the arms agree
within error bars.  Taken together,
the bandwidth budget and decoder configuration assumed for a predecoder may
need to be revisited under a more accurate noise model.

\section{Extended Related Work}
\label{sec:app-related}

\paragraph{Clifford and Pauli-noise baselines.}
Tableau simulation is the Clifford foundation \cite{Aaronson_Gottesman_2004}.
Stim pairs inverse-tableau references with SIMD Pauli-frame sampling
\cite{Gidney_2021}; STABSim and cuQuantum provide GPU Clifford kernels
\cite{Garner_Liu_Wang_Stein_Li_2025,Bayraktar_cuquantum_2023}.  \quasistim{}
does not replace these systems: it keeps Pauli-frame propagation as the
execution primitive and differs only in using a signed stabilizer-channel
representation.

\paragraph{QPD-based simulation.}
Pashayan et al.\ related estimator variance to negativity
\cite{Pashayan_Wallman_Bartlett_2015}.  Bennink et al.\ specialized this to
stabilizer-channel decompositions
\cite{Bennink_Ferragut_Humble_Laska_Nutaro_Pleszkoch_Pooser_2017}, our sampling
baseline; Hakkaku et al.\ and LeBlond et al.\ treat double decompositions of
unitary channels with Pauli noise
\cite{Hakkaku_Mitarai_Fujii_2021,LeBlond_Groszkowski_Lietz_Seck_Bennink_2025}.
These works establish the estimator.  \quasistim{} contributes reference
precomputation and batched execution so heterogeneous QPD patterns share
bit-sliced Pauli-frame operations.  Separate work reduces negativity before
sampling
\cite{Piveteau_Sutter_Woerner_2022,Kulikov_Yashin_Fedorov_Kiktenko_2024,Koukoulekidis_Kwon_Jee_Jennings_Kim_2022};
\quasifold{} specializes that line to measurement-heavy QEC by fusing
common-frame $R_Z$ axes, pruning rotations that cannot affect a later
measurement, and folding Pauli noise only when the site one-norm decreases.
Stratified sampling is complementary: it reallocates trajectories without
replacing the stabilizer backend
\cite{Dai_Koczor_2026,Myers_Teo_Mishra_Chai_Ng_2025}.

\paragraph{Alternative near-Clifford representations.}
When non-Clifford structure is compact globally, low-rank stabilizer
decompositions, active-subspace methods (Clifft), ZX compilation (Tsim),
generalized stabilizers (SOFT), and stabilizer tensor networks are preferable
\cite{Bravyi_Browne_Calpin_Campbell_Gosset_Howard_2019,Chase_Labib_2026,Haenel_Luo_Zhao_2026,Li_Zheng_Zhang_Lou_Ying_Liu_Sun_2025,MasotLlima_GarciaSaez_2024}.
They should outperform local QPD when the coherent resource stays globally
small.  \quasistim{} targets the opposite: many weak non-Pauli sites in a large
Clifford circuit.  Sparse and perturbative methods target selected observables
\cite{Rall_Liang_Cook_Kretschmer_2019,Rudolph_Jones_Teng_Angrisani_Holmes_2025,Begusic_Hejazi_Chan_2025,Tuloup_Ayral_2026,Miller_Ostrove_Hines_BlumeKohout_Young_Proctor_2025};
\deltaframe{} instead returns complete measurement records, which a decoder
requires.

\paragraph{Noisy trajectories and reuse.}
Prefix-sharing and analog-trajectory methods amortize expensive state updates
across similar paths
\cite{Granet_Hémery_Dreyer_2025,Wang_Tannu_Nair_2025,Dangwal_Oberoi_Sailopal_Shah_Chong_2025,Patti_Nguyen_Lietz_McCaskey_Khailany_2025}.
Neither condition holds here: frame updates are cheap, and complete QPD
patterns are nearly unique.

\paragraph{Rare-event simulation.}
Fail-fast and dynamical subset sampling reduce the trajectories needed to
observe logical failure
\cite{Beverland_Carroll_Cross_Yoder_2025,Heussen_Winter_Rispler_Muller_2024}.
They change which trajectories are drawn, not how each is executed, and are
complementary to \quasistim{}.

\section{\quasifold{} Representation and Rewrite Conditions}
\label{sec:app-qf-algebra}

\subsection{Site Tables and Coefficient Composition}

Each site stores a sign-free binary axis key and an orientation sign for
the fusion sum $\Theta_{\mathcal{G}}=\sum_{j\in\mathcal{G}}s_j\theta_j$. Measurements and resets advance the block
identifier; fusion candidates must share both key and block.

For a fixed axis $A$, the signed table is indexed by the axis monoid
\begin{equation}
  \mathcal{M}_A=\bigl\{Q B^k:Q\in\mathcal{P}_n,\ k\in\{0,1\}\bigr\}.
  \label{eq:axis-monoid}
\end{equation}
Here products denote composition of stabilizer channels, and global Pauli
phases are ignored. For a square-root-axis branch,
\begin{equation}
  B=\sqrt{A},\qquad B^2=A,\qquad QB=B^{\sigma(Q)}Q,
  \label{eq:sqrt-relations}
\end{equation}
where $\sigma(Q)=+1$ for a commuting Pauli and $-1$ otherwise. These relations
reduce products back to $k\in\{0,1\}$; coefficients with the same index are
added. This is an indexing convention, not a claim that the overcomplete
channel representation is unique.

A reset branch has a different algebra. In its physical frame, $B=R_q$ resets
qubit $q$, and
\begin{equation}
  B^2=B,\qquad BQ_q=B,
  \label{eq:reset-relations}
\end{equation}
for a Pauli supported on that qubit. A local reset does not erase Paulis on
other qubits: if $Q=Q_q\otimes Q_{\bar q}$, then
$BQ=(I_q\otimes Q_{\bar q})B$. Conversely, $Q_qB=B$ only when $Q_q$ fixes the
reset state. These identities are conjugated with the Clifford frame. The
implementation restricts damping folds to the reset's physical qubit; coherent
folds may use a larger support cap.

A rotation about $A$ has weights $a$ on $I$, $b$ on $A$, and $|\sin\Theta|$
on a square-root branch, with
\begin{equation}
  \begin{aligned}
    a&=\tfrac12(1+\cos\Theta-|\sin\Theta|),\\
    b&=\tfrac12(1-\cos\Theta-|\sin\Theta|).
  \end{aligned}
  \label{eq:rotation-table}
\end{equation}
The branch is $B$ when $\sin\Theta\geq0$ and $B^{-1}=AB$ otherwise. Thus the
branch weight stays nonnegative for arbitrary real angles, and
$\gamma(R_A(\Theta))=|\cos\Theta|+|\sin\Theta|$. Damping uses the three-term
decomposition in Appendix~\ref{sec:app-thermal-qpd}; a Pauli channel uses only
$k=0$ and has $\gamma=1$.

Composition multiplies coefficients and combines equal indices after reducing
branch products. The triangle inequality gives
$\gamma(\mathcal{E}_2\circ\mathcal{E}_1)
\leq\gamma(\mathcal{E}_1)\,\gamma(\mathcal{E}_2)$. For a normalized table with negative weight
$\eta$, $\gamma=1+2\eta$, so either quantity can serve as the cost test.

\subsection{Transparency Checks}
\label{sec:app-qf-transport}

Common-frame keys make Cliffords transparent. Other sites need the checks
below; measurements, resets, composites, and blocking channels remain
barriers. Timing markers are transparent. Rotations with commuting axes and
pairs of Pauli channels are transparent.

For a Pauli channel $\mathcal{N}(\rho)=\sum_Qp_QQ\rho Q$, commuting past a
rotation channel $\mathcal{R}_A$ requires covariance:
\begin{equation}
  \mathcal{N}\circ\mathcal{R}_A
  =\mathcal{R}_A\circ\mathcal{N}.
  \label{eq:channel-covariance}
\end{equation}
The implementation checks covariance for the rotation family about $A$ by
requiring $p_Q=p_{QA}$ for every Pauli $Q$ anticommuting with $A$, with missing
keys treated as zero. Stored coefficients are compared directly, not with a
commutation tolerance: a small anisotropy is still a different channel. This
angle-independent test is conservative for special angles that may admit
weaker conditions.

Damping requires each nonzero Pauli branch to commute with the common-frame
axis, fixing the reset state in its local frame. A rotation cannot commute past a reset.
This branchwise condition is stronger than the channel-level
covariance test used for rotations.

\subsection{Fusion Cost and Negativity Cancellation}

For a group with common axis $P$ along a transparent path, the rotation product is
$\prod_{j\in\mathcal G}R_{s_jP}(\theta_j)=R_P(\Theta_{\mathcal G})$.
The angle sum is $\Theta_{\mathcal{G}}=\sum_{j\in\mathcal{G}}s_j\theta_j$. Its cost is
non-increasing for arbitrary real angles, since
\begin{equation}
  \begin{aligned}
    \gamma(\alpha+\beta)
      &=|\cos(\alpha+\beta)|+|\sin(\alpha+\beta)|\\
      &\leq (|\cos\alpha|+|\sin\alpha|)
             (|\cos\beta|+|\sin\beta|)\\
      &=\gamma(\alpha)\gamma(\beta).
  \end{aligned}
  \label{eq:rotation-gain}
\end{equation}
The inequality follows by expanding the angle sums and applying the triangle
inequality. Strictness depends on the angles: distinct nonzero Clifford angles
can give equality. Fusion can still remove a redundant QPD site in that case.

Pauli folding provides a second source of cancellation. For a chunk supported
on $\{I,A\}$ with weight $p_A$ on $A$, the site's Pauli coefficients become
\begin{equation}
  \begin{aligned}
    a'&=a(1-p_A)+bp_A,\\
    b'&=ap_A+b(1-p_A).
  \end{aligned}
  \label{eq:fold-pauli-coefficients}
\end{equation}
For $a\geq0$ and $b\leq0$, $b'$ becomes nonnegative when
$ap_A\geq|b|(1-p_A)$. All negativity is removed only if the other resulting
coefficients are nonnegative as well, including $a'$. The resulting
$\gamma=1$ denotes a probability distribution over stabilizer branches; it does
not mean that the Pauli terms have disappeared. For larger chunks, convolution
can move negativity among several Pauli keys, so the pass evaluates the full
composed table rather than relying on this two-term example.

\section{\quasifold{} Search and Lowering}
\label{sec:app-qf-search}

Algorithm~\ref{alg:quasifold} gives the preprocessing order. Fusion and
observability decisions are planned against the input stream before any group
is rewritten. Fusion runs before folding because it can delete entire sites,
and a folded composite is no longer a plain rotation.

\begin{algorithm}[t]
\caption{\quasifold{} preprocessing. Candidate groups and chunks must satisfy
transparency and support constraints.}
\label{alg:quasifold}
\small
\begin{algorithmic}[1]
\Require circuit $\mathcal C$
\State $\mathcal S\gets\textsc{Scan}(\mathcal C)$
\State Plan same-axis fusion along transparent paths within each block
\State Plan observability pruning on the input stream
\State Apply the fusion and pruning plans
\If{angle canonicalization inserted a Clifford}
  \State Replay and re-scan to update the common frame
\EndIf
\ForAll{surviving axis sites $i$}
  \Repeat
    \State Accept eligible single chunks that lower $\gamma_i$
    \State Re-scan until no single chunk improves $\gamma_i$
    \State Try eligible pairs from a bounded candidate set
    \State Commit the best improving pair, if any
  \Until{neither phase improves $\gamma_i$}
\EndFor
\State \Return $\textsc{Replay}(\mathcal C,\mathcal S)$
\end{algorithmic}
\end{algorithm}

\paragraph{Angle canonicalization.}
A fused angle outside $[0,\pi/2]$ is split into a Clifford quarter turn and
a supported residual rotation. The new Clifford changes later axis keys, so
this case requires replay and re-scanning before folding; the usual path needs
one scan. Unrepresentable rewrites are not committed.

\paragraph{Candidate selection.}
Before- and after-noise distributions are accumulated separately. Eligible
chunks share the site's block, satisfy transparency and support constraints, and
are consumed at most once. Improving singletons are applied until reaching a fixed point,
then the best improving pair from a bounded reservoir is committed. The phases
repeat because a pair may expose new singleton gains. By default,
\texttt{pair\_lookahead} allows 32 candidates; values below two disable pairs.

A numerical threshold rejects negligible improvements; covariance uses exact
coefficient equality. This deterministic search can miss pairs outside the
reservoir and larger jointly beneficial groups. An external optimizer may
replace candidate evaluation while retaining transparency and replay checks
(\S\ref{sec:limitations}).

\paragraph{Lowering and replay.}
Tables return to the physical frame as local channels or supported multi-qubit
coherent composites. Damping folds remain single-qubit. Replay replaces sites
and removes consumed chunks without changing measurement order or Stim
metadata; the resulting circuit is passed to the sampler.

\section{\deltaframe{} Implementation and Cost Model}
\label{sec:app-deltaframe}

\subsection{Reference Tables}

Let $L$ be the number of measurement bits. The all-zero pattern fixes a
noiseless reference $r_0\in\mathbb F_2^L$, and each nonzero site digit defines
the single-site difference $\Delta_i(a)=r(i{=}a)\oplus r_0$. For two active
sites, define $\Delta_{ij}(a,b)=r(i{=}a,j{=}b)\oplus r_0$. A conflict is
recorded when
\begin{equation}
  \Delta_{ij}(a,b)\neq\Delta_i(a)\oplus\Delta_j(b).
  \label{eq:pair-conflict}
\end{equation}
The ordinary full site-pair sweep probes digit $1$ at each endpoint; the optional
hybrid mode checks further digit combinations until a conflict is found. A
restricted sweep considers only pairs in the same measurement slice. The graph
therefore depends on the selected probes, not just on the circuit.
Detected conflicts form edges in a graph over sites. For each connected
component $\mathcal B$, the table enumerates local configurations and stores
their measurement-bit differences from $r_0$. Joint entries include
higher-order effects within that block. The stored differences are sparse
index lists; a lane's reference is assembled as
$\widehat{r}(p)=r_0\oplus\bigoplus_{\mathcal{B}}\Delta_{\mathcal{B}}(p_{\mathcal{B}})$.

Additive offsets exploit the affine structure of stabilizer records, but
pairwise probes need not discover every higher-order dependency. Missed
inter-block effects can invalidate the prediction despite exact local table
entries. Section~\ref{sec:eval-accuracy} checks this empirically.

\paragraph{Measured block structure.}
Table~\ref{tab:block-structure} reports the conflict graph each evaluation
circuit produces under the full site-pair sweep. Two regimes appear. Thermal
relaxation records no conflict at any distance, so every site is its own block
and the reference is a plain XOR of single-site offsets; this is what makes the
restricted sweep sound for that family. Coherent noise does produce conflicts,
but the components stay small: the largest block has five sites on the logical
Hadamard circuits and four sites on every other circuit, well inside the enumeration limit of
\S\ref{sec:app-deltaframe-cost}. The defect-qubit configurations are the least
sparse: only two of $50$ and two of $98$ sites remain singletons at $d=5$ and
$d=7$, because concentrating rotations on a few data qubits makes their
influence cones overlap. Cost stays low there because the blocks are small,
not because they are few.

The last column gives what that structure costs. Precomputation is a one-time
charge of under a second on every circuit but the largest, and it is dominated
by the pair sweep rather than by block enumeration: at surface $d=11$ under
damping, $2{,}892$ sites take $5.49$\,s, of which $5.42$\,s is the sweep and
$0.06$\,s the enumeration. Enabling hybrid mode changes that picture, and the
added cost is motif training rather than the sweep: the transversal Hadamard
goes from $0.04$ to $1.25$\,s at $d=3$ and from $0.20$ to $20.6$\,s at $d=5$.
The block decomposition it produces is identical either way on these circuits,
so the wider sweep buys motifs rather than a finer partition.

\begin{table*}[t]
  \centering
  \small
  \caption{Conflict-block structure per evaluation circuit, and the wall time
  of reference precomputation on $128$ workers with hybrid mode disabled
  (median of three runs, timing \texttt{compile()} alone, so circuit
  construction and the \quasifold{} merge are excluded). Sizes are listed as
  $s^{\times n}$ for $n$ blocks of $s$ sites; the largest block in each row is
  the last entry. Coherent noise is $R_Z(0.006\pi)$; damping is
  $\tau/T_1=10^{-2}$.}
  \label{tab:block-structure}
  \begin{tabular}{@{}llrrrlr@{}}
    \toprule
    Circuit & Config. & Sites & Blocks & Conflicts & Block sizes & Precomp. (s) \\
    \midrule
    SC mem. & $d=3$, 2q & 51 & 34 & 25 & $1^{\times 23}$, $2^{\times 7}$, $3^{\times 2}$, $4^{\times 2}$ & 0.02 \\
    SC mem. & $d=5$, defects & 50 & 18 & 56 & $1^{\times 2}$, $2^{\times 8}$, $4^{\times 8}$ & 0.03 \\
    SC mem. & $d=7$, defects & 98 & 32 & 120 & $1^{\times 2}$, $2^{\times 12}$, $4^{\times 18}$ & 0.09 \\
    BB mem. & $d=6$, defects & 120 & 110 & 10 & $1^{\times 100}$, $2^{\times 10}$ & 0.14 \\
    MSC 1+2 & $d_1=3$ & 74 & 55 & 26 & $1^{\times 42}$, $2^{\times 8}$, $3^{\times 4}$, $4^{\times 1}$ & 0.03 \\
    MSC 1+2 & $d_1=5$ & 465 & 380 & 117 & $1^{\times 322}$, $2^{\times 36}$, $3^{\times 17}$, $4^{\times 5}$ & 0.82 \\
    Logical $H$ & $d=3$ & 69 & 46 & 41 & $1^{\times 35}$, $2^{\times 4}$, $3^{\times 3}$, $4^{\times 3}$, $5^{\times 1}$ & 0.04 \\
    Logical $H$ & $d=5$ & 230 & 160 & 116 & $1^{\times 122}$, $2^{\times 17}$, $3^{\times 13}$, $4^{\times 5}$, $5^{\times 3}$ & 0.20 \\
    Logical $S$ & merge & 192 & 129 & 102 & $1^{\times 97}$, $2^{\times 9}$, $3^{\times 15}$, $4^{\times 8}$ & 0.16 \\
    Logical $S$ & merge, $d=5$ & 564 & 409 & 244 & $1^{\times 327}$, $2^{\times 25}$, $3^{\times 41}$, $4^{\times 16}$ & 1.43 \\
    \midrule
    SC mem. & $d=3$, damp & 68 & 68 & 0 & $1^{\times 68}$ & 0.02 \\
    SC mem. & $d=5$, damp & 294 & 294 & 0 & $1^{\times 294}$ & 0.06 \\
    SC mem. & $d=7$, damp & 776 & 776 & 0 & $1^{\times 776}$ & 0.33 \\
    SC mem. & $d=9$, damp & 1{,}610 & 1{,}610 & 0 & $1^{\times 1610}$ & 1.74 \\
    SC mem. & $d=11$, damp & 2{,}892 & 2{,}892 & 0 & $1^{\times 2892}$ & 5.49 \\
    BB mem. & $d=6$, damp & 1{,}008 & 1{,}008 & 0 & $1^{\times 1008}$ & 0.81 \\
    \bottomrule
  \end{tabular}
\end{table*}

\paragraph{Comparison with exact-pattern caching.}
\label{sec:pattern-batching}
There are $|\Omega|=\prod_i(1+k_i)$ complete patterns. Grouping identical
patterns amortizes a tableau only when those patterns recur. Small branch
probabilities reduce the number of active sites, but a weight-$w$ pattern still
has $\binom Mw$ possible supports before branch digits are counted. This limits
reuse when a circuit contains many weak sites. Blockwise precomputation instead
enumerates $\sum_{\mathcal B}(\prod_{i\in\mathcal B}(1+k_i)-1)$ nonzero
local configurations, which is smaller when interactions remain confined to
small blocks.

\subsection{Masked Propagation and Signed Weights}

Each site draws a digit for each lane and forms a bit mask for every branch.
The grouped Pauli part is sampled conditionally on digit $0$; nonzero branch
operations act only on their own masks. Background Pauli noise remains on the
shared frame path. Thus distinct patterns share an operation stream, not an
identical sequence of active gates.

A Pauli changes the corresponding $x$ or $z$ frame bits under the mask. A
single-qubit Clifford applies its binary map to $(x,z)\in\mathbb F_2^2$.
Resets require a separate, nonunitary update. For a $Z$-basis reset under lane
mask $m$, the wordwise update is
\begin{equation}
  \begin{aligned}
    x&\leftarrow x\mathbin{\&}\neg m,\\
    z&\leftarrow (z\mathbin{\&}\neg m)
           \mathbin{|}(u\mathbin{\&}m),
  \end{aligned}
  \label{eq:masked-reset}
\end{equation}
where $u$ contains fresh random bits. Outside the mask, the old frame is
preserved. Inside it, one plane is cleared and the other randomized, as in
Stim's reset semantics. Clearing both planes would incorrectly remove the
randomness of a later anticommuting measurement. The analogous $X$ and $Y$
updates use their corresponding reset bases.

Negative Pauli terms toggle a sign ancilla, whose final measurement supplies
the QPD sign. Sampling uses absolute coefficients and the common weight
magnitude $\Gamma$. Block-reference offsets are XORed with the frame record;
the sign is passed separately to the weighted estimator.

\subsection{Cost and Enumeration Limits}
\label{sec:app-deltaframe-cost}

For $N$ shots at batch width $B_s$, let $G$ denote shared frame work and $D$
offset work. With $M$ sites and $L$ measurement bits, sampling without hybrid
fallback costs
\begin{equation}
  T_{\mathrm{sample}}=
  O\!\left(\left\lceil\frac{N}{B_s}\right\rceil(G+M+D+L)\right).
  \label{eq:deltaframe-sample-cost}
\end{equation}
This fixes the batch representation and branch alphabet, including SIMD word
work per batch; it does not imply unlimited speedup as $B_s$ grows.

For a full pair sweep, let $K_{\mathcal B}=\prod_{i\in\mathcal B}(1+k_i)$.
With bounded $k_i$ and tableau cost $T_{\mathrm{ref}}$, precomputation costs
\begin{equation}
  T_{\mathrm{pre}}=O\!\left(
    \left[M+M^2+\sum_{\mathcal B}(K_{\mathcal B}-1)\right]
    T_{\mathrm{ref}}\right).
  \label{eq:deltaframe-compile-cost}
\end{equation}
The terms count single-site runs, pair probes, and joint configurations.
For an unrestricted branch alphabet and exhaustive digit-pair probing, the
first two counts become $\sum_i k_i$ and $\sum_{i<j}k_i k_j$.

Compilation rejects a connected component with more than eight sites rather
than splitting it. Within an accepted block, all nonzero mixed-radix
configurations are enumerated, including those with some zero digits. There is
no separate configuration-count cap, so the branch alphabet also affects
memory and preprocessing cost.

\paragraph{Verification.}
Random draws check reference achievability, not equality with one canonical
record. Failed checks block sampling unless
\texttt{allow\_\allowbreak unverified} is set; this override does not repair
the model. Skipping the pair sweep or block-table construction also requires
it. The single-core timing benchmark uses both shortcuts with verification
disabled to isolate frame-execution cost, not verified precomputation.

The optional hybrid mode retains at most 1{,}024 inter-block corrections and 512
fallback motifs, truncating coverage at these limits. Retained motifs trigger
exact reference evaluation for affected lanes, sharing identical patterns
within a batch. This finite coverage is not a higher-order guarantee, and the
reference runs add costs beyond Eq.~\eqref{eq:deltaframe-sample-cost}.

\end{document}